\documentclass[aps,prb,twocolumn,floats,amsmath,amssymb,superscriptaddress,noeprint,nolongbibliography]{revtex4-2}

\usepackage{here}
\usepackage{subcaption}
\usepackage{tabularx}
\usepackage{graphicx}
\usepackage{bm}
\usepackage{dcolumn}
\usepackage{amsthm}
\usepackage{tikz}
\usepackage{siunitx}
\usepackage{hyperref}
\usepackage{booktabs}
\usepackage{multirow}
\usepackage{dcolumn}
\usepackage{blindtext}
\hypersetup{
  colorlinks,
  citecolor=blue,
  linkcolor=blue,
  urlcolor=blue}

\newcommand{\exciting}{{\usefont{T1}{lmtt}{b}{n}exciting}}
\newcommand{\te}[1]{{\text{#1}}}
\newcommand{\fett}[1]{\bm{\textbf{#1}}}
\newcommand{\mc}[1]{\multicolumn{1}{c}{#1}}

\newcommand{\eg}{{\it e.g.}, }
\newcommand{\ie}{{\it i.e.}, }

\begin{document}

\title{X-ray absorption spectroscopy of oligothiophene crystals from many-body perturbation theory}

\author{Konstantin Lion}
\affiliation{Institute of Physics and IRIS Adlershof, Humboldt-Universit\"at zu Berlin, Berlin, Germany}
\author{Caterina Cocchi} \thanks{Current address: Institute of Physics and Center for Nanoscale Dynamics (CeNaD), Carl von Ossietzky Universit\"at Oldenburg, Oldenburg, Germany}
\affiliation{Institute of Physics and IRIS Adlershof, Humboldt-Universit\"at zu Berlin, Berlin, Germany}
\author{Claudia Draxl}
\affiliation{Institute of Physics and IRIS Adlershof, Humboldt-Universit\"at zu Berlin, Berlin, Germany}

\date{\today}

\begin{abstract}
We present an x-ray absorption spectroscopy study from the carbon $K$, sulfur $K$, and sulfur $L_{2,3}$ edges of crystalline oligothiophenes of varying length, \ie bithiophene~(2T), quaterthiophene~(4T), and sexithiophene~(6T), performed from first principles by means of all-electron density-functional theory and many-body perturbation theory. A comprehensive assignment of all relevant spectral features is performed based on the electronic structure and the character of the target conduction states. The inclusion of electron-hole effects leads to significant redistribution of oscillator strengths and to strongly bound excitons with binding energies ranging from \SIrange{1.5}{4.5}{\electronvolt}. When going from 2T to 6T, exciton binding energies decrease by up to \SI{1}{\electronvolt}, which we attribute to the reduction of the average Coulomb attraction with increasing oligomer length.  These high values are significantly larger than their counterparts in the optical excitations of these systems and indicative of their localization on the respective molecules. For the same reason, local-field effects which typically dominate the optical absorption of organic crystals, turn out to play only a negligible role at all edges. We identify two sets of carbon atoms, \ie with or without sulfur bonding, which exhibit distinct features at the C $K$-edge. The sulfur atoms, on the other hand, yield similar contributions in the S, $K$, and $L_{2,3}$ edge spectra. Our results show excellent agreement with available experimental data.
\end{abstract}
\maketitle

\section{Introduction}

With the aim of producing cheap devices on a large scale, much effort has been devoted to identify potential active components in electronics and optoelectronics. In this context, organic materials based on $\pi$-conjugated molecules have emerged as outstanding candidates for organic field-effect transistors~(OFETs)~\cite{Horowitz1998,akimichi1991,liu2022,chen2020}, organic solar cells~(OSCs)~\cite{Granstrom1998,yuan2019,sun2022,zheng2022}, and organic light-emitting diodes~(OLEDs)~\cite{Forrest2004,FORREST2003,salehi2019,bauri2021} due to their strong light-matter interaction in the visible range of the solar spectrum and their low molecular weight. Among them, oligo-~and polythiophenes offer the unique combination of chemical stability, efficient electronic conjugation, and synthetic flexibility which allows for the deliberate adjustment of properties through substitution at the thiophene ring~\cite{handbook2}. Poly~(3-hexylthiophene)~(P3HT) has already established itself as an organic semiconductor for OFETs and OSCs~\cite{1P3HT,2P3HT,holliday2016,yang2020,chatterjee2021}. Oligothiophenes bear the advantage of having a well defined structure, and therefore produce more defect-free thin films compared to polythiophenes. Among others, $\alpha$-sexithiophene is a very promising candidate for the use in OFETs~\cite{FET_6T}. Optimizing the performance of these materials requires extensive knowledge of their chemical composition and fundamental properties, including electronic structure and their response to electromagnetic radiation.

The carbon $K$ absorption edge of molecular systems is commonly investigated, \eg to determine the orientation of the molecules on a substrate~\cite{stohr_NEXAFS}. Among the $\pi$-conjugated molecular crystals, oligothiophenes have been rather well studied in this context. High-temperature~(HT)~bithiophene~(2T) and quaterthiophene~(4T) monolayers on metal surfaces are typically oriented with their molecular planes parallel to the substrate~\cite{stohr_thiophene_pt,thiophene_CK,Agthiophene} while sexithiophene~(6T) adopts a more upright geometry on glass substrates~\cite{Opitz,pithan2015} when forming thin films. Short chains up to terthiophene~(3T) in the gas phase have been studied~\cite{bithiophene_CK_theory}, whereas the spectral features of longer oligomers are less explored. Additionally, the absorption from the C $K$ edge of the thiophene monomer is well investigated but it is increasingly difficult to interpret corresponding spectra for longer oligomers due to the presence of a higher number of bands in the underlying electronic structure.

The absorption from the sulfur $K$ edge is often used to study the chemical composition of sulfur-containing fossil fuels. In this context, monothiophene~\cite{thiophene_CK,perera,mijovilovich}, substituted thiophenes~\cite{george}, and aromatic thiophenic compounds~\cite{thioether,mijovilovich} have been investigated. To the best of our knowledge, results for longer oligothiophene chains are still missing. The sulfur $L_{2,3}$ edge has been studied experimentally for different oligothiophene films such as monothiophene~\cite{thiophene_CK,stohr_thiophene_pt,Agthiophene,baseggio2017}, bithiophene~\cite{Agthiophene,Koller_SL23,Ramsey_SL23,baseggio2017}, and polythiophene films~\cite{polythiophene}. The main goal was to identify the formation of chemisorptive bonds with a substrate, and it revealed the cleavage of the C-S bond of monothiophene films on Pt(111)~\cite{stohr_thiophene_pt}. The assignment of the spectral features, however, appears rather controversial in the literature~\cite{thiophene_CK,Agthiophene,Ramsey_SL23,Koller_SL23,baseggio2017}. 

On the theory side, first-principles studies of core spectra supplementing experimental results are often performed in the (half)-core-hole approximation. While this approach is known to be accurate for absorption from the $K$ edge, where spin-orbit coupling (SOC) effects, usually disregarded in these calculations, are typically negligible,  $L_{2,3}$-edge spectra can hardly be reproduced. The Bethe-Salpeter equation (BSE), however, employed in this work enables us to accurately treat SOC and to obtain not only reliable spectra but also full insight into the nature of the core-level excitations \cite{vorwerk2,vorwerk_2019}. Based on this approach, we investigate the x-ray absorption spectra of oligothiophene crystals of different length (termed nT, where n indicates the number of monomer units) from the C and S $K$-edge, as well as from the S $L_{2,3}$ edge, providing a comprehensive assignment of the spectral features and an in-depth analysis of their origin in terms of electronic contributions, also with respect to the oligomer length.   

\section{Theoretical background}
X-ray absorption spectra~(XAS) are obtained from first principles through the solution of the Bethe-Salpeter equation of many-body perturbation theory~(MBPT)~\cite{Strinati1988}, which can be mapped onto an effective eigenvalue problem
\begin{equation}
\sum_{c' u' \fett{k}'}H^{BSE}_{c u \fett{k}, c' u' \fett{k}'} A^{\lambda}_{c' u' \fett{k}'} = E^{\lambda} A^{\lambda}_{c u \fett{k}}, \label{eq:BSE}
\end{equation}
where $c$ and $u$ denote the initial core states and the final unoccupied states, respectively. The Hamiltonian in Eq.~\eqref{eq:BSE} can be split into three contributions:
\begin{equation}
H^{BSE} = H^{diag} + H^{x} + H^{dir}. \label{eq:BSE_parts}
\end{equation}
The diagonal term, $H^{diag}$, describes single-particle transitions; solely including this term corresponds to the independent-particle approximation (IPA). The exchange term, $H^{x}$, reflects the repulsive bare Coulomb interaction, while the direct term, $H^{dir}$, contains the attractive screened Coulomb interaction. The eigenvalues, $E^{\lambda}$ in Eq.~\eqref{eq:BSE}, represent excitation energies and their resonances in the absorption spectra. Here, we define exciton binding energies as the difference between excitation energies calculated from the IPA and the BSE, respectively, \ie $E_{b} = E_{\text{IPA}}^{\lambda} - E_{\text{BSE}}^{\lambda}$. The absorption spectrum is expressed by the imaginary part of the macroscopic dielectric tensor,
\begin{equation}
\te{Im} \,\epsilon_{\te{M}}(\omega) = \frac{8 \pi^2}{\Omega}\left| \fett{t}_{\lambda}\right|^2 \delta(\omega - E_{\lambda}) .\label{eq:Im}
\end{equation}
The BSE eigenvectors $A^{\lambda}_{c u \fett{k}}$ determine the electron-hole (e-h) wavefunctions
\begin{equation}
\Phi^{\lambda}(\fett{r}_{e}, \fett{r}_{h}) = \sum_{c u \fett{k}} A^{\lambda}_{c u \fett{k}} \psi_{u \fett{k}}(\fett{r}_e) \psi^{*}_{c \fett{k}}(\fett{r}_h)  \label{eq:WF}
\end{equation}
and enter Eq.~\eqref{eq:Im} through the transition coefficients
\begin{equation}
\fett{t}_{\lambda} = \sum_{c u \fett{k}} A^{\lambda}_{c u \fett{k}} \frac{\left\langle c\fett{k}\left| \hat{\fett{p}}\right|u\fett{k}\right\rangle}{\varepsilon_{u \fett{k}} - \varepsilon_{c \fett{k}}}.
\end{equation}
In XAS, the BSE Hamiltonian can be furthermore separated into atomic contributions featuring the atom-selective character of the core-level excitations. The imaginary part of the macroscopic tensor can therefore be expressed as a sum over the contributions from the individual atomic species $\gamma$,
\begin{align}
\te{Im}\,\epsilon_{\te{M}} = \sum_{\gamma }\te{Im}\,\epsilon_{\te{M}}^{\gamma} . \label{eq:splitBSE}
\end{align}
This allows us to analyze the site-dependence of such excitations at carbon and sulfur species individually. 

\section{Computational details \label{sec:1}}
All calculations are performed using the full-potential all-electron code \exciting\ \cite{Exciting}. Treating valence and core electrons on equal footing, \exciting\ allows one to handle atomic species of any kind and study excitations from deep core levels to the shallow valence region. In the framework of the linearized augmented planewave plus local orbital (LAPW+lo) method, we treat the $1s$, $2s$, and $2p$ states of sulfur, and the $1s$ state of carbon as core states. XAS are calculated via the solution of the BSE with a fully relativistic treatment of core states \cite{vorwerk2}.

The Kohn-Sham electronic structure is computed within the local-density approximation (LDA) in the Perdew-Wang parametrization \cite{LDA}). For the groundstate calculations, we employ $\mathbf{k}$-grids of $8\times 8 \times 6$ for 2T, $3\times 5 \times 2$ for 4T, and  $3\times 5\times 1$ for 6T, respectively. The muffin-tin radii $R_{\te{MT}}$ are chosen to be $\SI{1.2}{\bohr}$ for C, $\SI{0.8}{\bohr}$ for H, and $\SI{2.0}{\bohr}$ for S. A planewave cut-off of $R_{\te{MT}}^{\te{min}}\,G_{\te{max}}=5.0$ is used for all systems, where $R_{\te{MT}}^{\te{min}}$ refers to the smallest muffin-tin sphere, \ie that of hydrogen. 

Quasiparticle energies are approximated by the Kohn-Sham eigenvalues, and thus, we expect the absorption onset to be underestimated in the order of $\SI{10}{\electronvolt}$. A scissors operator is therefore applied to align the calculated spectra to experimental references when available ($\SI{24.2}{\electronvolt}$ for the 2T carbon $K$-edge~\cite{Agthiophene} and $\SI{15.3}{\electronvolt}$ for the 2T sulfur $L_{2,3}$-edge~\cite{Agthiophene}). This is common practice for XAS computed from the BSE~\cite{vorwerk2,azobenzene,cocchi2018_2}. The screening entering the expression of the Coulomb potential is calculated in the random phase approximation, including all valence bands and 200 unoccupied states for all absorption edges.
The computational parameters used for the calculation of the different absorption edges are summarized in the Appendix. We checked that they ensure a convergence of the spectral shape and an accuracy of \SI{20}{\milli\electronvolt} for the lowest excitation energy. 

Core excitations typically exhibit ultrashort lifetimes and, therefore, large intrinsic broadenings that increase with the depth of the absorption edge. In lack of information on the lifetimes of individual excitations, we choose not to apply an energy-dependent broadening~\cite{L23_broadening} but employ a Lorentzian broadening of \SI{150}{\milli\electronvolt} (if not specified otherwise) that allows us to analyze all spectral features.

All input and output files are available on NOMAD~\cite{draxl2019} at the following link: \url{http://doi.org/10.17172/NOMAD/2023.03.30-1} 

\section{Crystal structures}

\begin{figure} \centering
\captionsetup[subfigure]{font=large,labelfont=large}
\begin{subfigure}{0.5\textwidth}
\caption{}
\includegraphics[width=0.95\linewidth]{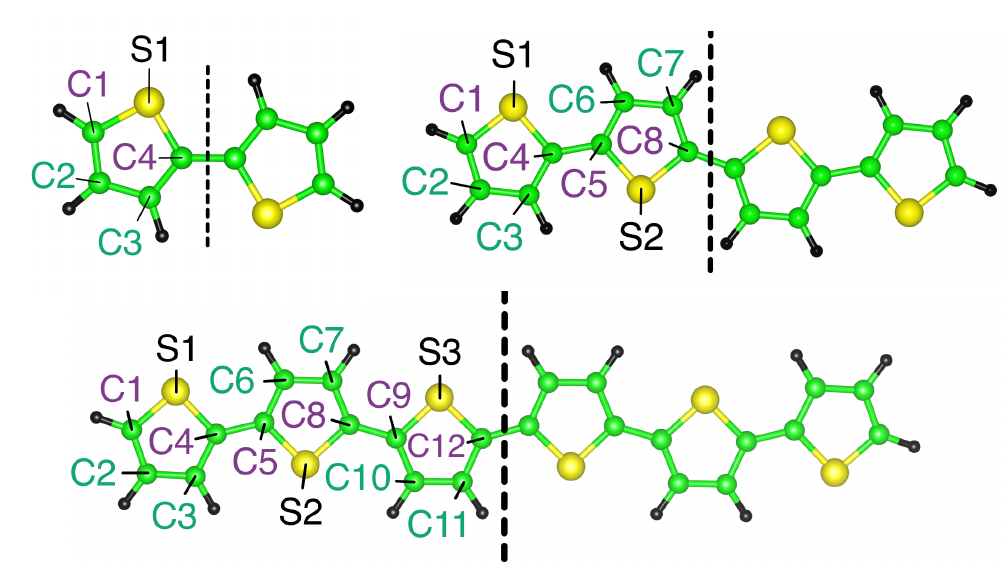}
\label{fig:labelling}
\end{subfigure}
\begin{subfigure}{0.2\textwidth}
\caption{}
\includegraphics[width=0.9\linewidth]{Pictures/2T_molecule_orbitals_new.pdf}
\label{fig:orbitals}
\end{subfigure}
\quad
\begin{subfigure}{0.25\textwidth}
\caption{}
\includegraphics[width=0.9\linewidth]{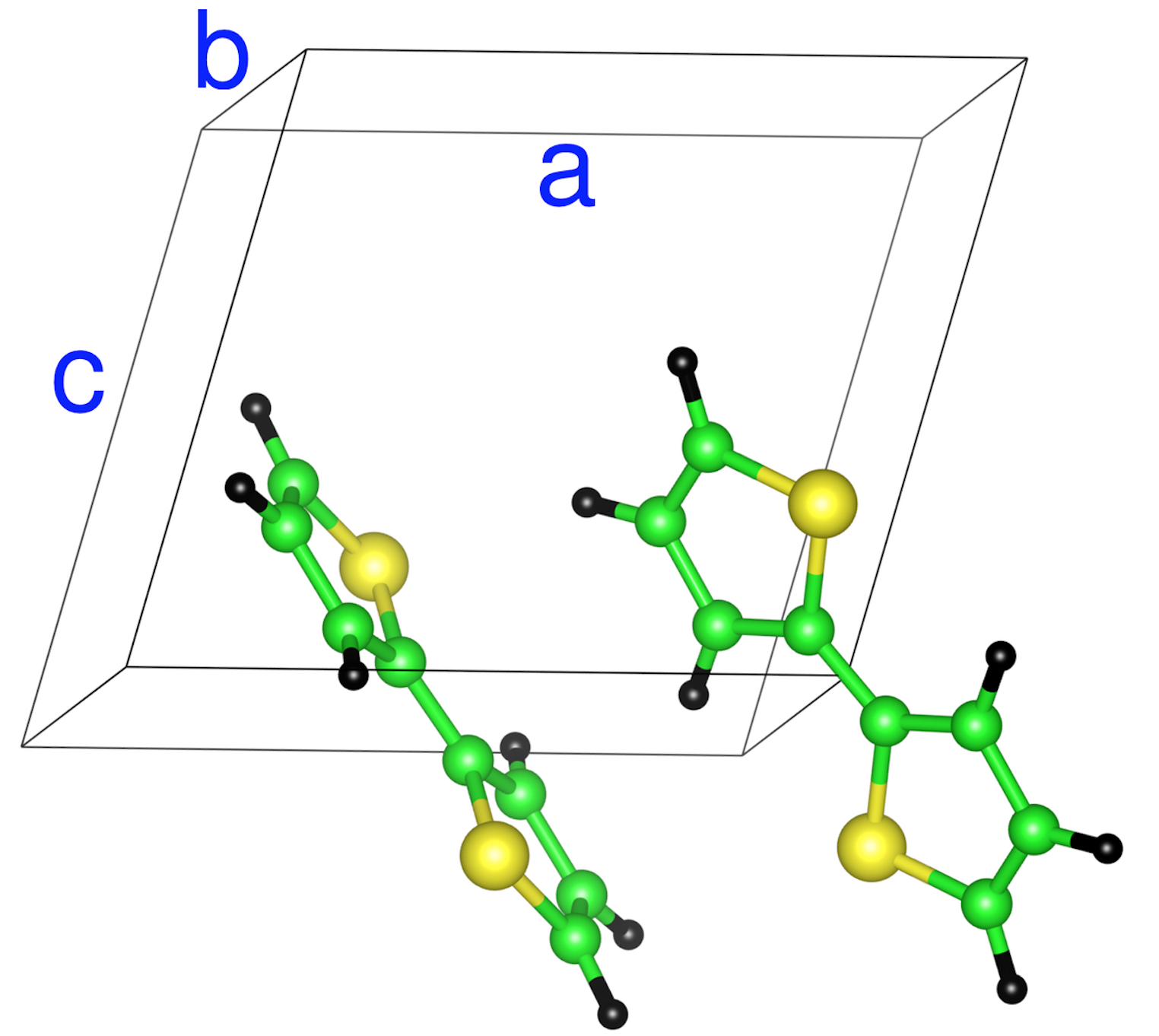}
\label{fig:crystal}
\end{subfigure}
\caption{(a) 2T, 4T, and 6T oligomers, showing the nomenclature of the inequivalent carbon and sulfur atoms. The black dotted lines indicate the reflection symmetry of the respective oligomer.  Carbon atoms are given in green, sulfur atoms in yellow, and hydrogen atoms in black. (b) Sketch of the $\pi^{*}$ and $\sigma^{*}$ orbitals in the 2T oligomer, taken as representative of the nT series considered here. (c) Unit cell of the 2T crystal with lattice parameters $a$, $b$, and $c$ built by two inequivalent molecules in the typical herringbone arrangement. 
\label{fig:nTstructure}}
\end{figure}

\begin{table}
\caption{Chemical formula and lattice parameters of 2T, 4T, and 6T crystals, adopted from Refs. \onlinecite{2T_structure}, \onlinecite{4T_structure}, and \onlinecite{6T_Klett}, respectively.\label{tab:lattice_para}}
\begin{ruledtabular}
\begin{tabular}{lcdddd}
 & Formula &\mc{$a$ {[\si{\angstrom}]}} & \mc{$b$ {[\si{\angstrom}]}} & \mc{$c$ {[\si{\angstrom}]}} & \mc{$\beta$ {[$^\circ$]}} \vspace{2pt}\\ 
 \hline \\[-8pt]
$\alpha$-2T    & C$_8$H$_8$S$_2$   &  8.81 & 5.77 & 7.87 & 107.1  \\
$\alpha$-4T/HT & C$_{16}$H$_{16}$S$_4$   &  8.93 & 5.75 & 14.34 & 97.2 \\
$\alpha$-6T/HT & C$_{24}$H$_{24}$S$_6$  & 9.14 & 5.68 & 20.67 & 97.8  \\
\end{tabular}
\end{ruledtabular}
\end{table}

The oligothiophene crystals considered in this work, are composed of two molecules per unit cell, where each molecule has the general formula unit n(C$_4$H$_2$S). We consider representatives of different lengths with an even number of rings, \ie 2T, 4T, and 6T. The carbon atoms in each molecule can be divided into two groups, \ie those with a covalent bond to sulfur (referred to as $\alpha$-C) and those without such a bond (referred to as $\beta$-C). The oligothiophene molecules are depicted in Fig.~\ref{fig:labelling}, also including the labeling adopted hereafter for the chemically inequivalent atoms. Each thiophene ring consists of $sp^2$ hybridized carbon and sulfur atoms, as well as hydrogen atoms saturating the dangling bonds. Here, we focus only on $\alpha$-nT~\cite{tour1992}, also known as 2,2'-nT, where the thiophene rings are connected at the $\alpha$-C sites, \eg C1 and C4 in 2T, see Fig.~\ref{fig:labelling}. In these aromatic heterocyclic molecules, the spatial orientation of the $\sigma^*$ orbitals can be represented by the plane spanned by the atoms and the $\pi^*$ orbitals by a vector perpendicular to this plane. This is illustrated for the 2T molecule in Fig.~\ref{fig:orbitals}. The aromatic character of these molecules results in a (quasi) planar configuration, which is preserved in their crystalline form.

The nT crystal structure is characterized by the herringbone arrangement of the inequivalent molecules. Such an arrangement is commonly found in organic crystals consisting of planar linear molecular chains~\cite{hotta2004}. While there is only one polymorph of crystalline $\alpha$-2T, two polymorphs have been identified for the $\alpha$-4T and $\alpha$-6T crystals depending on the growth conditions: a low temperature phase with four inequivalent molecules and a high temperature~(HT) phase with two. We have chosen to focus solely on the HT phase in order to directly compare our results across different oligothiophenes. At ambient conditions, oligothiophenes crystallize in a monoclinic structure where $\alpha$-2T belongs to the space group $P$2$_1$/$c$ whereas $\alpha$-4T/HT and $\alpha$-6T/HT exhibit space group $P$2$_1$/$a$. The lattice parameters and chemical formulas of the investigated structures are listed in Table~\ref{tab:lattice_para}. The unit cell of crystalline 2T is exemplarily shown in Fig.~\ref{fig:crystal}.

\section{Results}

\subsection{Electronic structure}
\begin{figure*}
\includegraphics[width=1.\linewidth]{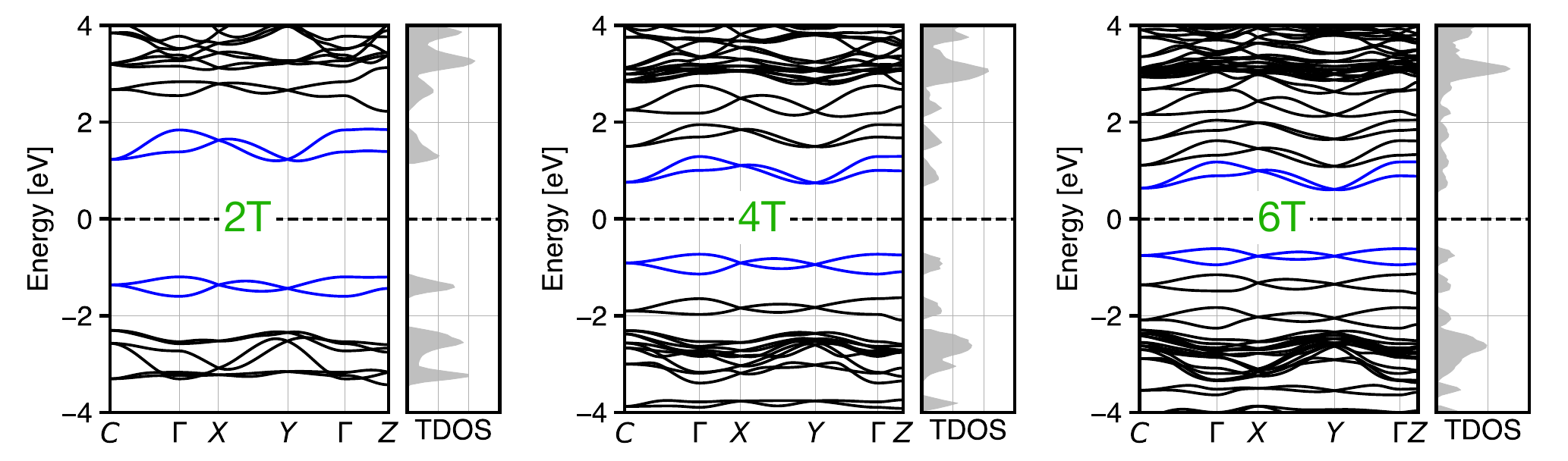}
\caption{Kohn-Sham band structure and total density of states (TDOS) of 2T (left), 4T (middle), and 6T (right). Energies are relative to the Fermi level set in the mid-gap. The subbands of the highest valence-band pair and the lowest conduction-band pair are highlighted in blue. The considered high-symmetry points in units of ($2\pi/a$, $2\pi/b$, $2\pi/c$) are $\Gamma=(0,0,0)$, $\te{C}=(0.5,0.5,0)$, $\te{X}=(0.5,0,0)$, $\te{Y}=(0,0.5,0)$, and $\te{Z}=(0,0,0.5)$. \label{fig:BS_DOS}}
\end{figure*}

\begin{figure}
\includegraphics[width=1.0\linewidth]{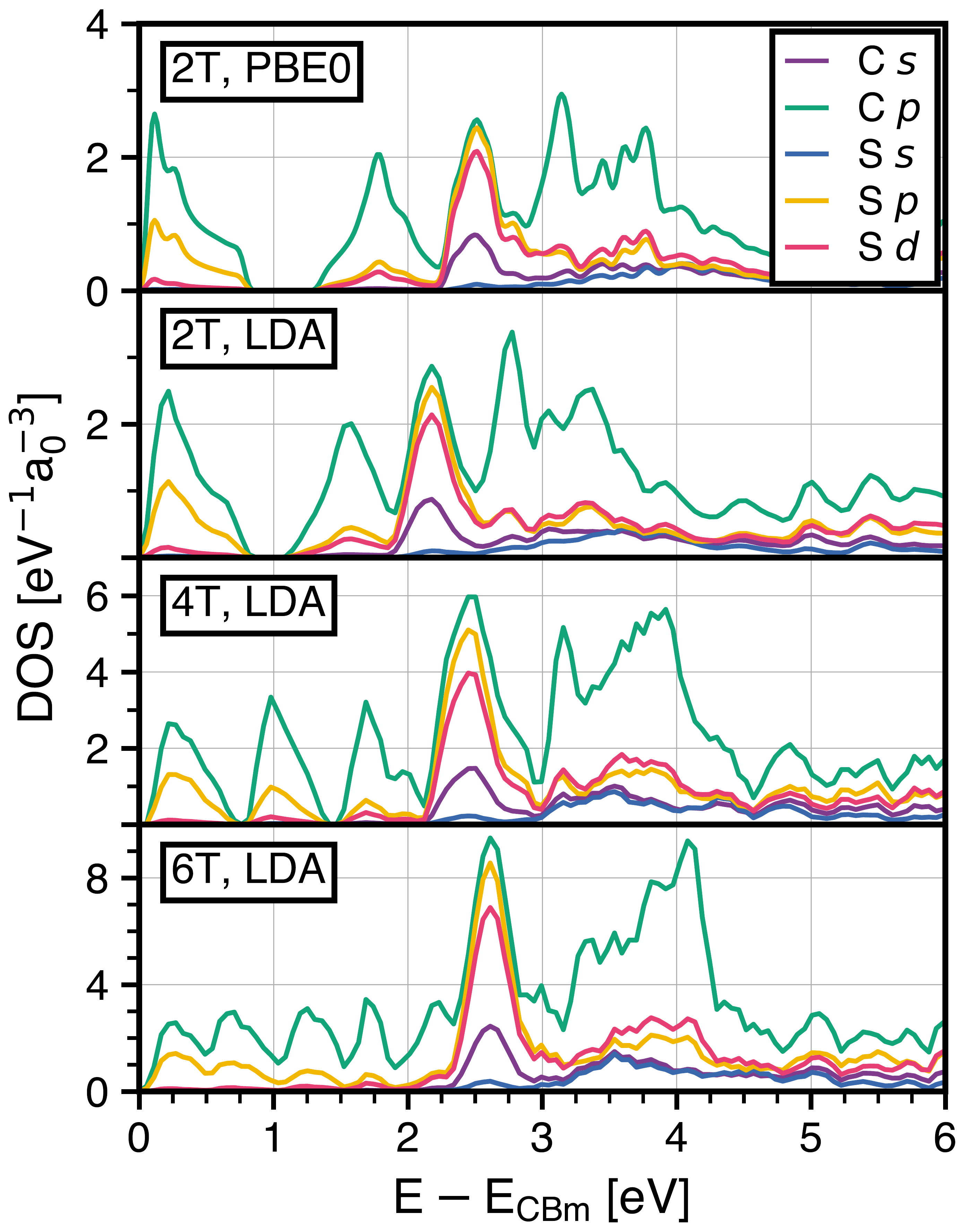}
\caption{Projected densities of states indicating hydrogen, carbon, and sulfur atomic orbital contributions in 2T (top), 4T (middle), and 6T  (bottom). Energies are relative to the conduction band minimum (CBm). Calculations are performed with the LDA or PBE0 functional. \label{fig:PDOS}}
\end{figure}

In the first step of our analysis, we investigate the electronic structure of the nT crystals. In Fig.~\ref{fig:BS_DOS}, the band structures (left panels) and the densities of states (DOS, right panels) of crystalline 2T, 4T, and 6T are depicted for an energy region of $\pm$\SI{4}{\electronvolt} around the band gap. Our results for 4T and 6T show overall good agreement with previously published DFT results~\cite{6T_Klett,draxl_band_6T}. Qualitatively, semi-empirical extended H\"uckel theory ~\cite{4T_structure,6T_structure} provides a similar picture. A characteristic feature of molecular crystals is evident: Each band is split due to the presence of two inequivalent molecules in the unit cell. The lowest conduction-band pair (corresponding to the pair of the lowest unoccupied molecular orbitals, LUMO pair) and highest valence-band pair (corresponding to the pair of the highest occupied molecular orbitals, HOMO pair) are highlighted in blue. With increasing molecular length, and hence increasing number of electrons in the system, the number of bands in both the valence and conduction regions increases, also reflected in a higher DOS. The corresponding peaks in the lower conduction bands are well separated in 2T and 4T but are overlapping in 6T, forming an electronic continuum. 

The band pairs are generally non-degenerate, except at the Brillouin zone boundaries, X and Y, along the $\fett{a}^*$ and $\fett{b}^*$ directions, \ie the normal vectors w.r.t. the lattice parameters $a$ and $b$. The band splitting of the HOMO pair is maximal at the $\Gamma$ point, with values of \SI{454}{\milli\electronvolt}, \SI{286}{\milli\electronvolt}, and \SI{287}{\milli\electronvolt} for 2T, 4T, and 6T, respectively. Previously reported values of \SI{450}{\milli\electronvolt} for 4T and \SI{420}{\milli\electronvolt} for 6T \cite{4T_structure,6T_structure} obtained by a semi-empirical quantum-chemistry approach are higher than ours. The largest band dispersion for the HOMO pair is along  $\overline{\Gamma\te{C}}$ [related to the ($\fett{a}^*, \fett{b}^*$) plane], $\overline{\Gamma\te{X}}$, and $\overline{\Gamma\te{Y}}$ (parallel to the $\fett{a}^*$ and $\fett{b}^*$ axis, respectively). Minimal dispersion is found along $\overline{\Gamma\te{Z}}$, which being parallel to the $\fett{c}^*$ axis, represents approximately the long molecular axis. Charge-carrier mobilities are therefore expected to be highest in the ($\fett{a}^*, \fett{b}^*$) plane regardless of oligomer length. Similiar results have also been found in other molecular crystals, such as in oligoacenes~\cite{oligoacene_dispersion} and sexiphenyl~\cite{draxl_banddispersion}. The overall flat band character of the conduction states around the band gap can be attributed to the dominant role of the $\pi$ orbitals.

The projected density of states (PDOS) of the conduction bands is shown in Fig.~\ref{fig:PDOS}. The sharp peaks up to \SI{3}{\electronvolt} from the conduction-band edge have mainly C $p$ and S $p$ character and are formed by antibonding $\pi^*$ orbitals. They are therefore expected to participate significantly in the C $K$-edge and S $K$-edge absorption spectra. The contribution from the C $p$ states is similiar for all subbands below \SI{2}{\electronvolt}, while that of the S $p$ states is decreasing when going to higher energy. There are also small contributions from S $d$ states. This admixture of S $d$ states, while seemingly insignificant, plays a crucial role in explaining the aromatic character of oligothiophenes~\cite{Thiophene_orbitals} as well as their electronic structure~\cite{Thiophene_pd_hyb}. The LUMO subbands are therefore expected to contribute to the absorption from the S $L_{2,3}$ edge. 

The DOS of all investigated systems is largest at approximately \SIrange{2}{2.5}{\electronvolt} where multiple hybridized states contribute (see also Fig.~\ref{fig:BS_DOS}), including the $\sigma^*$(C-S), the $\sigma^*$(C-C), and the $\sigma^*$(C-H) orbitals. Consequently, we expect significant contributions from these bands to all investigated absorption edges. The region from \SIrange{3}{4.5}{\electronvolt} has mainly C $p$ character and can be attributed to higher-lying $\sigma^*$(C-C) orbitals with small contributions from S $d$, S $p$, S $s$, and C $s$ states. From \SIrange{5}{8}{\electronvolt}, we find hybridized states of S $p$, S $d$, and C $p$ character. The overall PDOS is very similiar for all investigated systems. The main differences are the occurrence of additional bands and a slight redshift of the strongest peak with increasing oligomer length. Additionally, from \SIrange{3}{4}{\electronvolt}, the contributions from C $p$ states shift to higher energies with increasing oligomer length. 

For 2T, we also show in Fig.~\ref{fig:PDOS} the PDOS of the unoccupied region obtained with the hybrid functional PBE0~\cite{ernzerhof1999}. The small differences compared to the LDA result justify to use LDA and apply a scissors shift for mimicking self-energy effects when computing the XAS spectra.

\subsection{X-ray absorption spectra \label{sec:xray_introduction}}
In the following, we will show our results for the x-ray absorption spectra from the $K$ and $L_{2,3}$ edges. A detailed analysis of the spectral features for the C $K$ and S $L_{2,3}$ edges is performed for 2T, where experimental data are available~\cite{Agthiophene}. In the case of the S $K$ edge, where this is not the case, we focus on 4T to highlight the differences between the two inequivalent sulfur sites. We then investigate the effects of oligomer length on the spectral features and exciton binding energies for all considered systems.

\begin{figure*}
\includegraphics[width=1.0\linewidth]{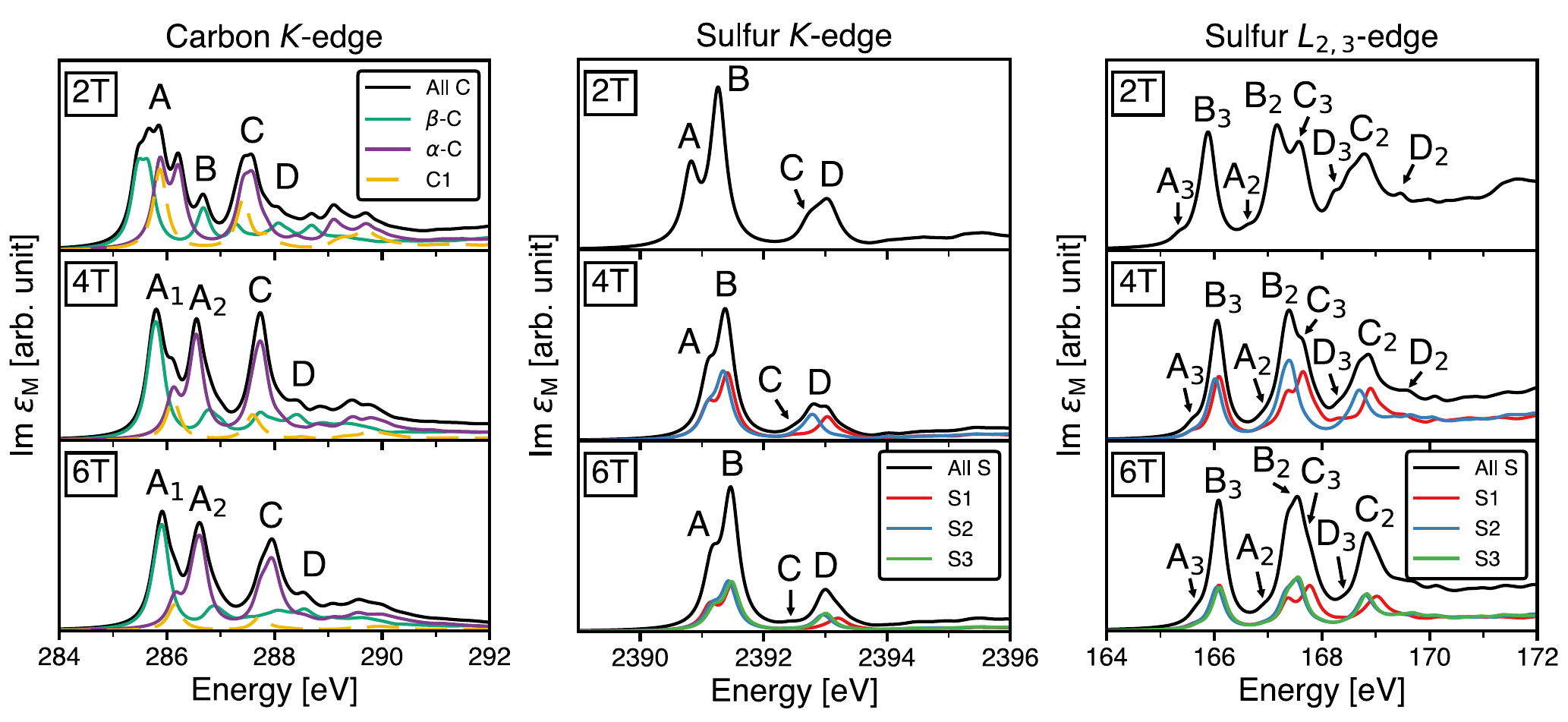}
\caption{Absorption spectra from the carbon $K$ (left), sulfur $K$ (middle), and sulfur $L_{2,3}$ (right) absorption edge of crystalline 2T~(top), 4T~(middle), and 6T~(bottom) as obtained from solutions of the BSE. The contributions from inequivalent atoms are shown as colored lines, their sum in black. All spectra are averages over the three diagonal cartesian components. For the assignment of spectral features, we refer to Tables~\ref{tab:CK_edge_peaks}, \ref{tab:SK_edge_peaks}, and \ref{tab:SL23_edge_peaks}.\label{fig:all_spectra_comparison}}
\end{figure*}

The absorption spectra of all investigated edges are shown in Fig.~\ref{fig:all_spectra_comparison}. The black lines represent the sum of contributions from all inequivalent atoms. To guide the reader, we label the peaks in two ways: first, by their ascending energy order in the full absorption spectrum, and second, by their origin. In the case of the sulfur S $L_{2,3}$ spectra, the subscripts 2 and 3 indicate whether the feature stems from the $L_2$ or $L_3$ edge, respectively. Our approach allows us not only to probe each atom individually, but also to clearly discern the nature of the excitations. A detailed explanation of all peaks and their origin is given in the respective sections.

\subsubsection{Carbon $K$ edge}

\begin{figure}
\includegraphics[width=0.95\linewidth]{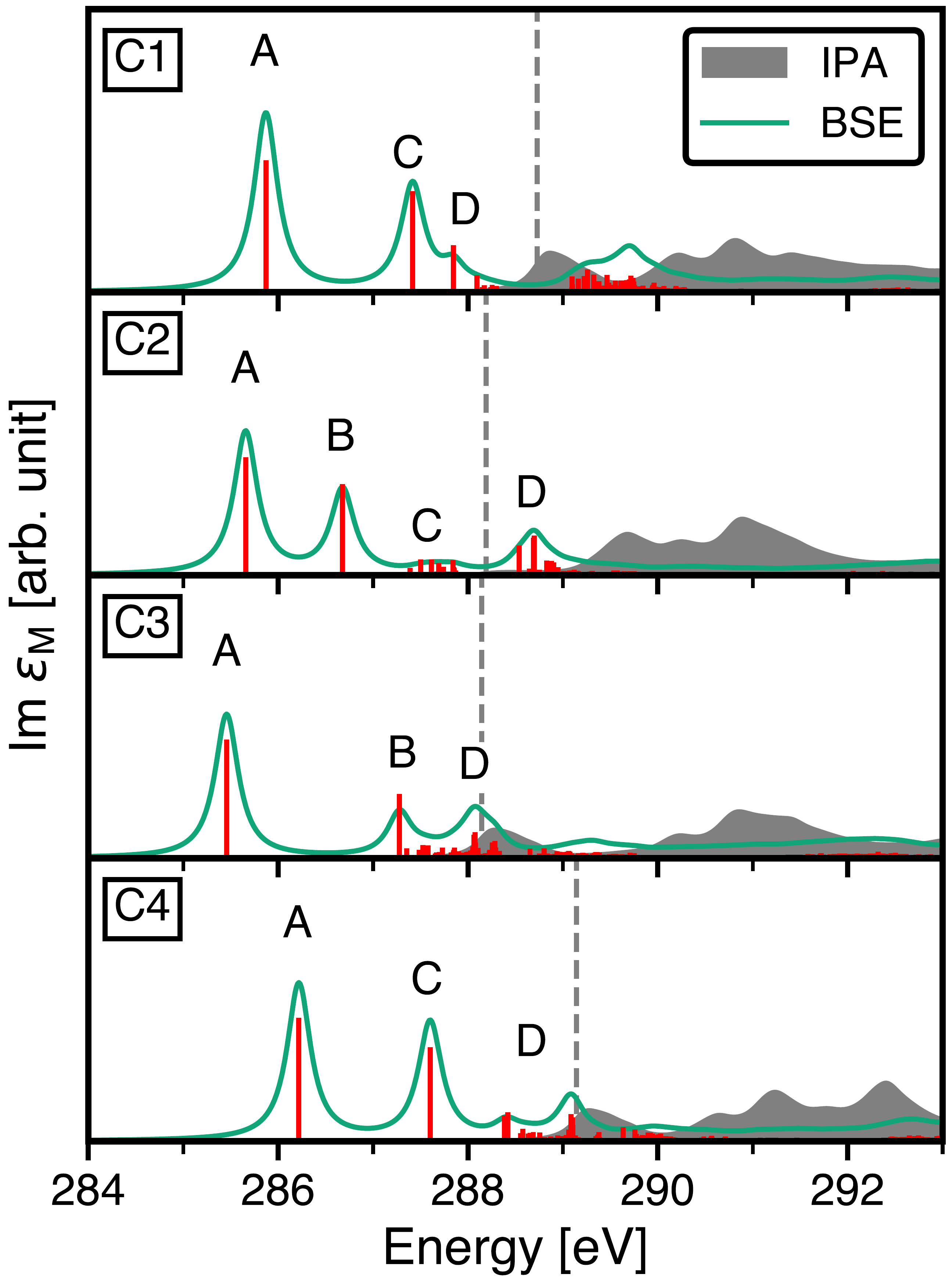}
\caption{$K$ edge absorption spectra of the inequivalent carbon atoms in 2T, averaged over the diagonal cartesian components. Excitation energies are indicated by the red bars. For comparison, the IPA results are shown (gray areas), the dashed bars mark the corresponding onset. 
\label{fig:2T_CK_comparison}}
\end{figure}

\begin{figure}
\includegraphics[width=1.\linewidth]{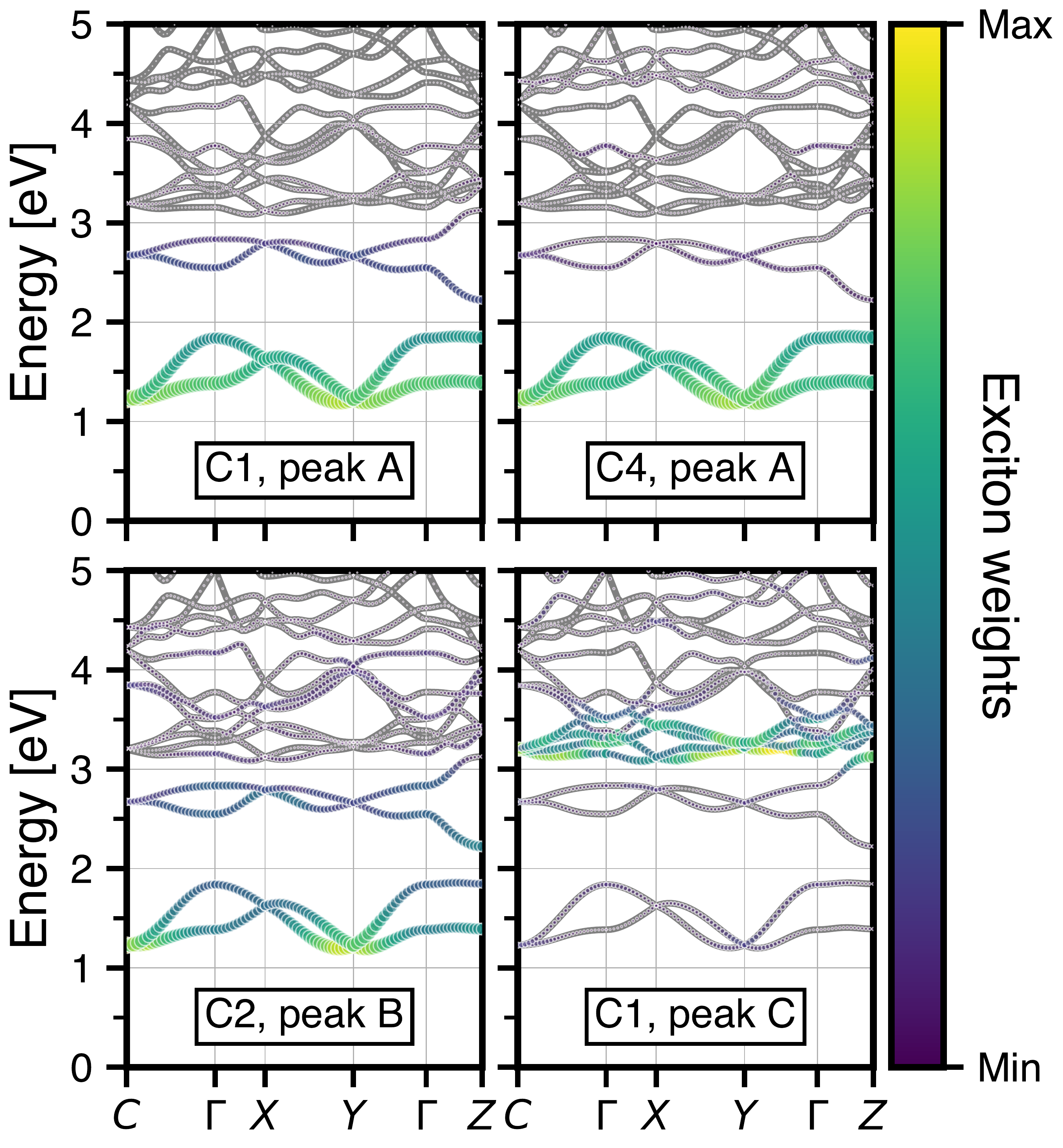}
\caption{Conduction band contributions to bound excitons with the largest oscillator strengths at the carbon $K$ edge of 2T. The size and color of the circles are indicative of the exciton weights.
\label{fig:2T_CK_excitons}}
\end{figure}

\begin{figure}
\includegraphics[width=1\linewidth]{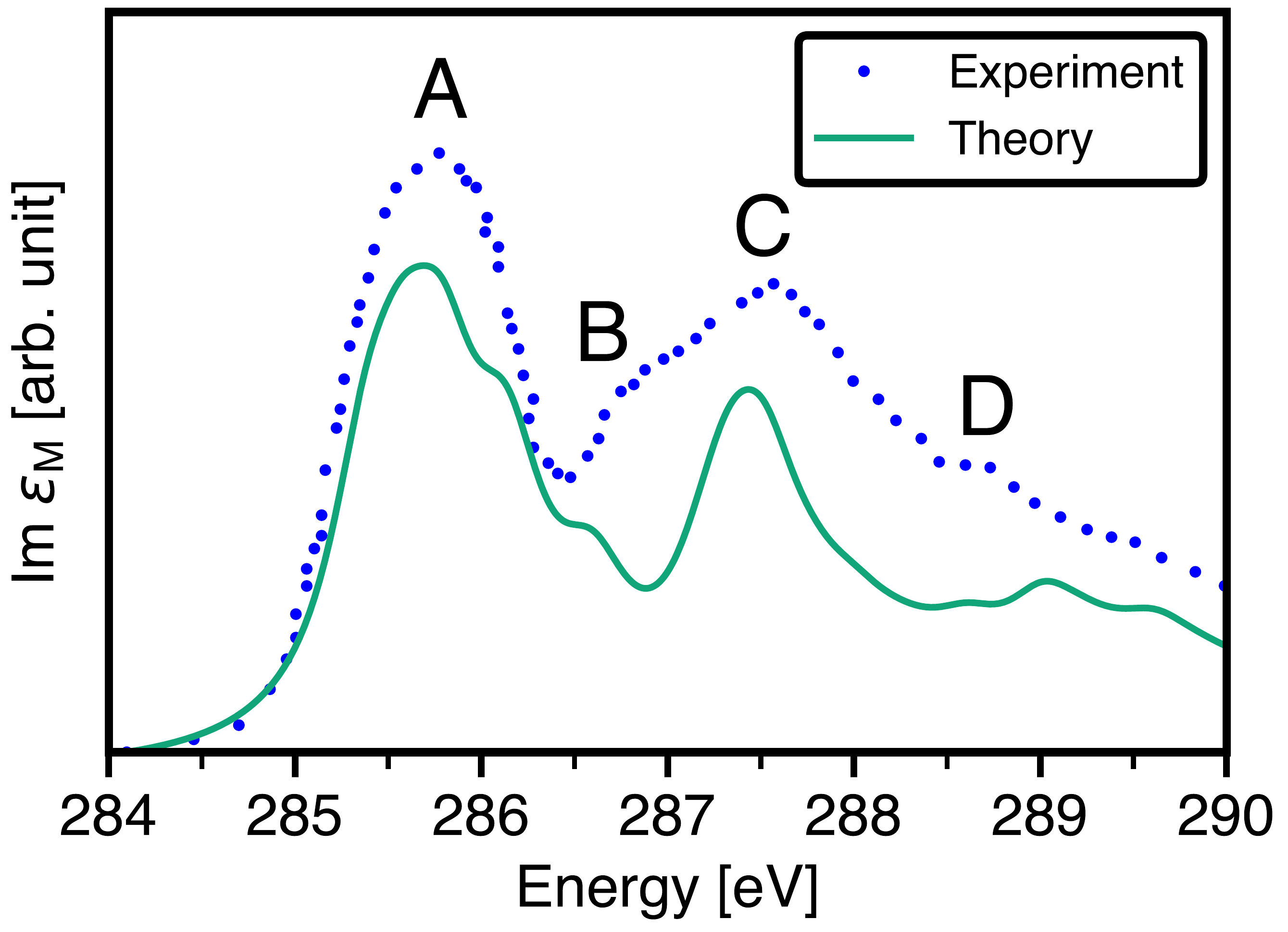}
\caption{Absorption spectra from the carbon $K$ edge of crystalline 2T including contributions from all inequivalent C atoms (as depicted in Fig. \ref{fig:2T_CK_comparison}). The calculated spectrum (green line) obtained by the BSE includes a Lorentzian broadening of \SI{250}{\milli\electronvolt}. It is shifted by \SI{24.2}{\electronvolt} to align it with the first absorption peak of the experimental reference (blue dots) taken from Ref. \cite{Agthiophene}.
\label{fig:2T_CK_exp_comparison}}
\end{figure}

\begin{figure*}
\includegraphics[width=0.9\linewidth]{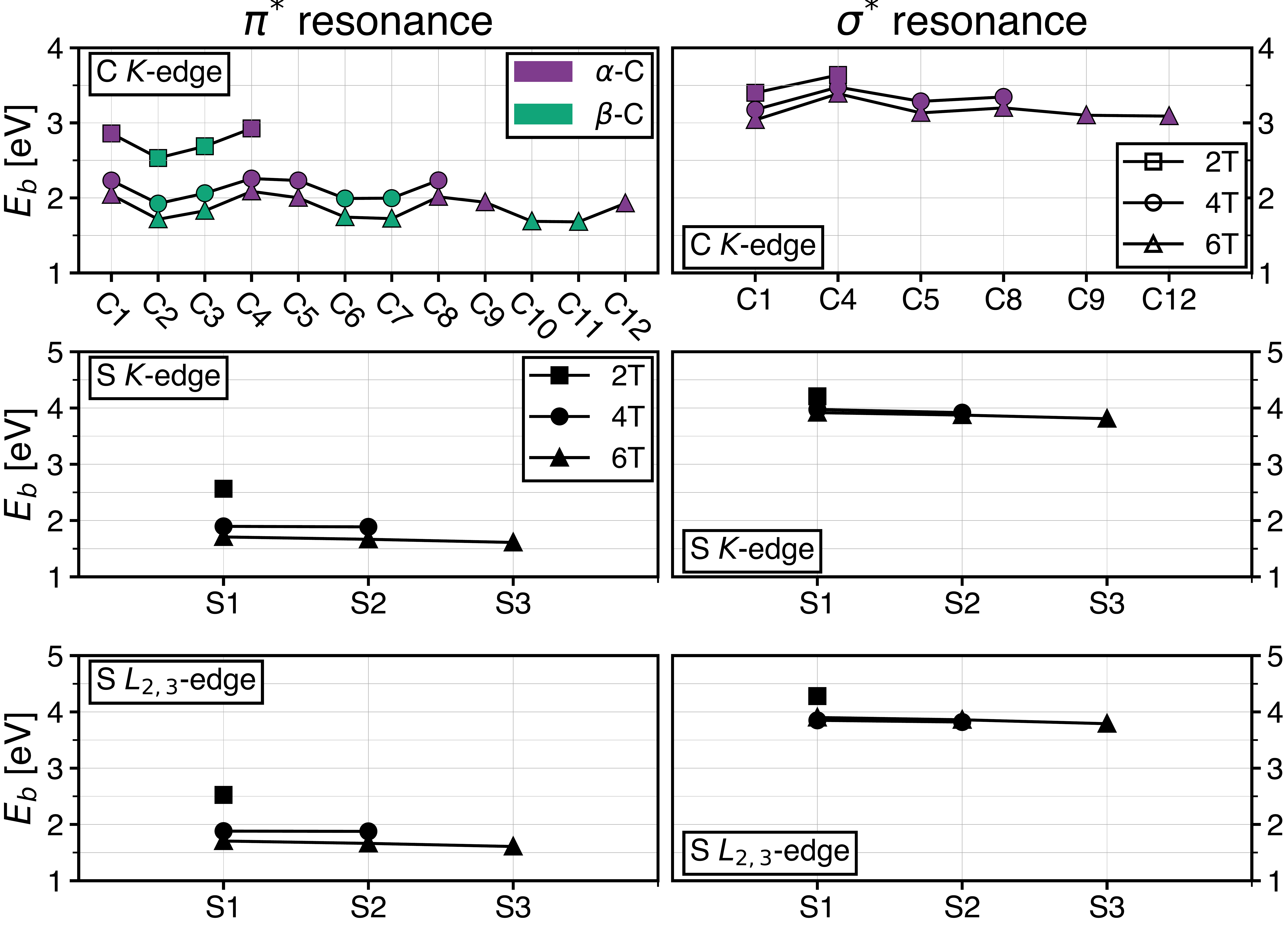}
\caption{Exciton binding energies ($E_b$) of all inequivalent atoms for the lowest-lying $\pi^{*}$ and $\sigma^{*}$ resonances in the carbon $K$, sulfur $K$, and sulfur $L_{2,3}$ absorption edges in crystalline nT ($n$=2,4,6). They are obtained as the difference between the excitation energies computed from the IPA and the BSE.
\label{fig:C_K_S_K_exciton_comparison}}
\end{figure*}

We start our analysis of the x-ray absorption spectra from the carbon $K$ edge by comparing the BSE solutions with the IPA for the 2T crystal. In Fig.~\ref{fig:2T_CK_comparison}, we show the spectra of the inequivalent C atoms as an average of the diagonal cartesian components, in order to reproduce the experimental scenario in which the samples have either polycrystalline domains or are randomly oriented with respect to the radiation source. We recall that the IPA spectra are related to the PDOS of the conduction states, featuring the contributions of the momentum matrix elements between core and conduction states, $\left\langle c\fett{k}\left| \hat{\fett{p}}\right|u\fett{k}\right\rangle$, \i.e. the dipole selection rules. Since only transitions from the 1$s$ state to unoccupied $p$ orbitals are dipole-allowed, the IPA spectra can be related to the corresponding contributions of the C $p$ states shown in Fig.~\ref{fig:PDOS}.

The first peak at the IPA onset represents transitions to the LUMO subbands. They are weak for C2 because the corresponding charge density is not localized on this atom. Beyond the first peak, a broader range of excitations of about \SIrange{3}{4}{\electronvolt} is found. They are formed by transitions to the conduction bands in the range from \SIrange{2}{5}{\electronvolt} above the onset (see Fig.~\ref{fig:PDOS}). Inclusion of the attractive electron-hole interaction by the BSE lowers the absorption onset by more than \SI{2.5}{\electronvolt} and leads to a significant redistribution of oscillator strength to a few excitons. As shown in Fig.~\ref{fig:all_spectra_comparison_LFE} in the Appendix, the differences between singlet and triplet excitation energies are smaller than \SI{150}{\milli\electronvolt}, indicating that local field effects (LFE) do not play a major role. This is a result of the highly localized character of the excitons, as previously found for the nitrogen $K$ edge of azobenzene monolayers~\cite{azobenzene}. This is in contrast to optical excitations of nT crystals, where we find singlet-triplet splitting of the same order as the binding energies~\cite{draft_thiophene}. 

We can identify several peaks in the BSE spectrum. A summary of the spectral features and their assignment is given in Table~\ref{tab:CK_edge_peaks}.The lowest excitation, peak A, is formed by transitions to the LUMO and LUMO+1 subbands with hybridized C-S $\pi^{*}$ character. This is visualized in Fig.~\ref{fig:2T_CK_excitons}, where we show for selected examples which bands contribute to the excitons with highest oscillator strength. Since all C atoms contribute to the aromaticity of 2T, this bright exciton with large oscillator strength is present in all atom-resolved spectra. The range of the respective excitation energies (about \SI{0.8}{\electronvolt}) corresponds to the different energies of the $1s$ core levels that are separated by \SI{1}{\electronvolt}. We also observe significant differences between $\alpha$-C and $\beta$-C species. The higher excitation energies of the $\alpha$-C is consistent with the higher electronegativity of sulfur compared to carbon. The highest excitation energy of peak A is found for the C4 atom which connects the two thiophene monomers without a bond to a hydrogen atom. 

Our results reproduce a trend that was previously observed for polycyclic aromatic hydrocarbons where C atoms bound to hydrogen have lower excitation energies than C atoms without such bonds~\cite{carbon_hydrogen}. Similiar findings were reported for 2T in the gas phase using the half core-hole approximation~\cite{bithiophene_CK_theory}. The $\beta$-C atoms, C2 and C3, also exhibit lower exciton binding energies of $E_{b}=\SI{2.66}{\electronvolt}$ and $E_{b}=\SI{2.88}{\electronvolt}$ compared to the $\alpha$-C atoms, C1 and C4, with $E_{b}=\SI{3.01}{\electronvolt}$ and $E_{b}=\SI{3.06}{\electronvolt}$. Such binding energies are typical for Frenkel excitons in molecular crystals~\cite{organic,azobenzene}. They are significantly larger than those in the optical excitations of oligothiophene crystals, which are typically below \SI{1}{\electronvolt}~\cite{draft_thiophene}. We also find that the two lowest-lying excitons -- one being bright, the other basically dark -- are very close in energy. These exciton splittings, which are related to the symmetry of the molecules, range from \SI{152}{\milli\electronvolt} for C1 to \SI{125}{\milli\electronvolt} for C3. 

The second peak, B, at \SI{286.7}{\electronvolt} originates from the $\beta$-C atoms and is dominated by an exciton with large oscillator strength. It is of $\pi^*$ character and formed by transitions to the LUMO and LUMO+1 subbands for C2, but mostly to the LUMO+1 subbands for C3. The third peak, C, at \SIrange{287.5}{287.6}{\electronvolt} is attributed to states associated with the C-S bond of the $\alpha$-C atoms. It is formed by transitions to $\sigma^*$ bands, ranging from \SIrange{3.0}{3.5}{\electronvolt} in the PDOS (Fig.~\ref{fig:PDOS}). Additionally, there are small contributions from delocalized excitons with $\sigma^*$ character, originating from C2. Peak D receives contributions from all carbon atoms where many transitions with mixed $\sigma^*$ and $\pi^*$ character occur. It is evident from this discussion that the chemical environment of the carbon atoms has a distinct impact on the spectral features.

\begin{table}
\caption{Excitation energies, $E$, of the spectral features in the carbon $K$ absorption edge of crystalline 2T and assignment of features to the respective final states. The excitation energies are defined for each inequivalent atom by the most intense exciton contributing to the respective peaks.\label{tab:CK_edge_peaks}}
\begin{ruledtabular}
\begin{tabular}{lcccc}
 {Peak / Assignment} & \multicolumn{4}{c}{$E$ [\si{\electronvolt}]} \\
 \cline{2-5}
 & C1 & C2 & C3 & C4 \\
 \hline \\[-8pt]
A / $\pi^*$(LUMO,LUMO+1)   & 285.9 & 285.7 & 285.5 & 286.2 \\
B / $\pi^*$(LUMO,LUMO+1)   & / & 286.7 & 287.3 & / \\
C / $\sigma^*$(C-S)        & 287.4 & 287.5 & / & 287.6 \\
D / mixed $\sigma^*$,$\pi^*$ & 287.8  & 288.7 & 288.1 & 288.4 \\
\end{tabular}
\end{ruledtabular}
\end{table}

Our theoretical results are in very good agreement with x-ray absorption measurement data for 2T multilayers on Ag(111)~\cite{Agthiophene}. Since no information on the experimental setup is available, we show in Fig.~\ref{fig:2T_CK_exp_comparison} the average over the diagonal cartesian components of the dielectric tensor. The individual components are shown in the Appendix (Fig~\ref{fig:all_spectra_comparison_crystalline}). We accurately reproduce the main spectral features shown in Fig.~\ref{fig:2T_CK_exp_comparison}, \ie (A) a broad resonance corresponding to transitions to the LUMO subbands, (B) a shoulder-like resonance due to transitions to the LUMO+1 subbands, (C) a third resonance due to transitions to higher bands with $\sigma^*$ character associated with the $\alpha$-C atoms, and (D) a shoulder assigned to higher Rydberg excitations~\cite{Agthiophene}. In our results, peak B appears as a shoulder of peak A instead of peak C. The relative energy of peak B with respect to the other spectral signatures, however, is well replicated. Due to the limited experimental resolution, it is not possible to identify all excitonic features contributing to peak D. The remaining peaks, however, are clearly resolved. Interestingly, the spectra for the bulk material discussed here compare very well to those of experimentally investigated multilayers~\cite{Agthiophene} which can be rationalized as follows: Since the intermolecular van der Waals interactions are significantly weaker than the covalent bonding within the molecules, the core excitations are strongly localized on the corresponding molecules, and thus, the spectral features are mainly determined by intramolecular interactions. 

We now explore the dependence of the spectra on the oligomer length. The overall spectra depicted in Fig.~\ref{fig:all_spectra_comparison}, left panels, show remarkably little differences. The spectral features of the four inequivalent carbon atoms of 2T can be clearly resolved. The longer oligomers, 4T and 6T, contain eight and twelve inequivalent carbon atoms, respectively. Summing over all contributions leads to smoother spectral shapes for these crystals compared to 2T. 

The most pronounced differences between the spectra occur close to the absorption onset in the range from \SIrange{284}{287}{\electronvolt}. Here, the $\pi^*$ resonances are blueshifted with increasing oligomer length, \ie the lowest lying excitation of 2T is shifted by \SI{0.24}{\electronvolt} and \SI{0.31}{\electronvolt} compared to its counterpart in 4T and 6T, respectively. In the latter two, the resonance A is split into two distinct peaks, A$_1$ and A$_2$, that are separated by \SI{0.9}{\electronvolt}, corresponding to the difference between 1$s$ core energies of chemically inequivalent carbon atoms giving rise to these excitations. Peak B is not resolved in the spectra of 4T and 6T but contributes to peak A$_2$ instead. We attribute the blueshift of the $\pi^*$ resonances to two effects that are commonly found in linear oligomers such as oligoacenes~\cite{Hummer_oligoacene_exciton}: With increasing oligomer length, the e-h pair is more delocalized, going hand in hand with increased dielectric screening. This reduces the average Coulomb attraction. The effect is more pronounced when going from 2T to 4T than from 4T to 6T because the e-h pair is still mainly localized on the respective atom and does not spread over the whole length of the molecule. It is important, however, to distinguish where the probed atom is exactly located in the molecule. The delocalization of the e-h pair decreases the closer the atom is to the edge of the molecule. This results in a shift to lower excitation energies for the atoms on the outer thiophene ring compared to those in the inner ones. Moreover, C1 experiences a significant redshift compared to the other $\alpha$-C atoms because of its additional hydrogen bond. This effect is illustrated in Fig.~\ref{fig:all_spectra_comparison} where the contributions stemming from the $\alpha$-~and the {$\beta$-C} atoms are depicted separately. In 4T and 6T, we can identify a shoulder between peaks A$_1$ and A$_2$ as originating from excitations of C1 with $\pi^*$ character. Peak C, however, is not significantly shifted for C1. With increasing oligomer length, we are therefore able to clearly distinguish the contributions of the C atoms based on their covalent bonding to S and H atoms. 

Comparing the binding energies of the lowest-lying excitons for all inequivalent C atoms (see Fig.~\ref{fig:C_K_S_K_exciton_comparison}), we find that binding energies related to transitions from the $\alpha$-C atoms are larger than those from the $\beta$-C atoms. We attribute this to the higher electronegativity of sulfur compared to carbon. With increasing oligomer length, the exciton binding energies of the lowest excitonic states slightly decrease for both carbon types. The splitting of the lowest-lying excitons reduces from \SI{150}{\milli\electronvolt} in 2T to \SI{50}{\milli\electronvolt} in 6T. This follows the trend observed for the band gap that depends almost linearly on the inverse molecular length~\cite{bandgap_conjugated}, and the exciton binding energies in the optical range of crystalline nT~\cite{draft_thiophene}. Again, we attribute this effect to the reduction of the average Coulomb interaction with increasing oligomer length as the excitons corresponding to $\pi^*$ resonances are increasingly delocalized along the molecular chain. We emphasize that for the lowest-lying exciton pair in all investigated systems, this delocalization does not give rise to charge transfer to adjacent molecules. This is in contrast to optical excitations where charge-transfer excitons can be found for long-chain molecular crystals~\cite{bussi2002,Hummer_oligoacene_exciton,hummer2004,puschnig2002,bechstedt,sharifzadeh2013,cocchi2018}.

\subsubsection{Sulfur $K$ edge \label{sec:S_K_edge}}

\begin{figure}
\includegraphics[width=0.95\linewidth]{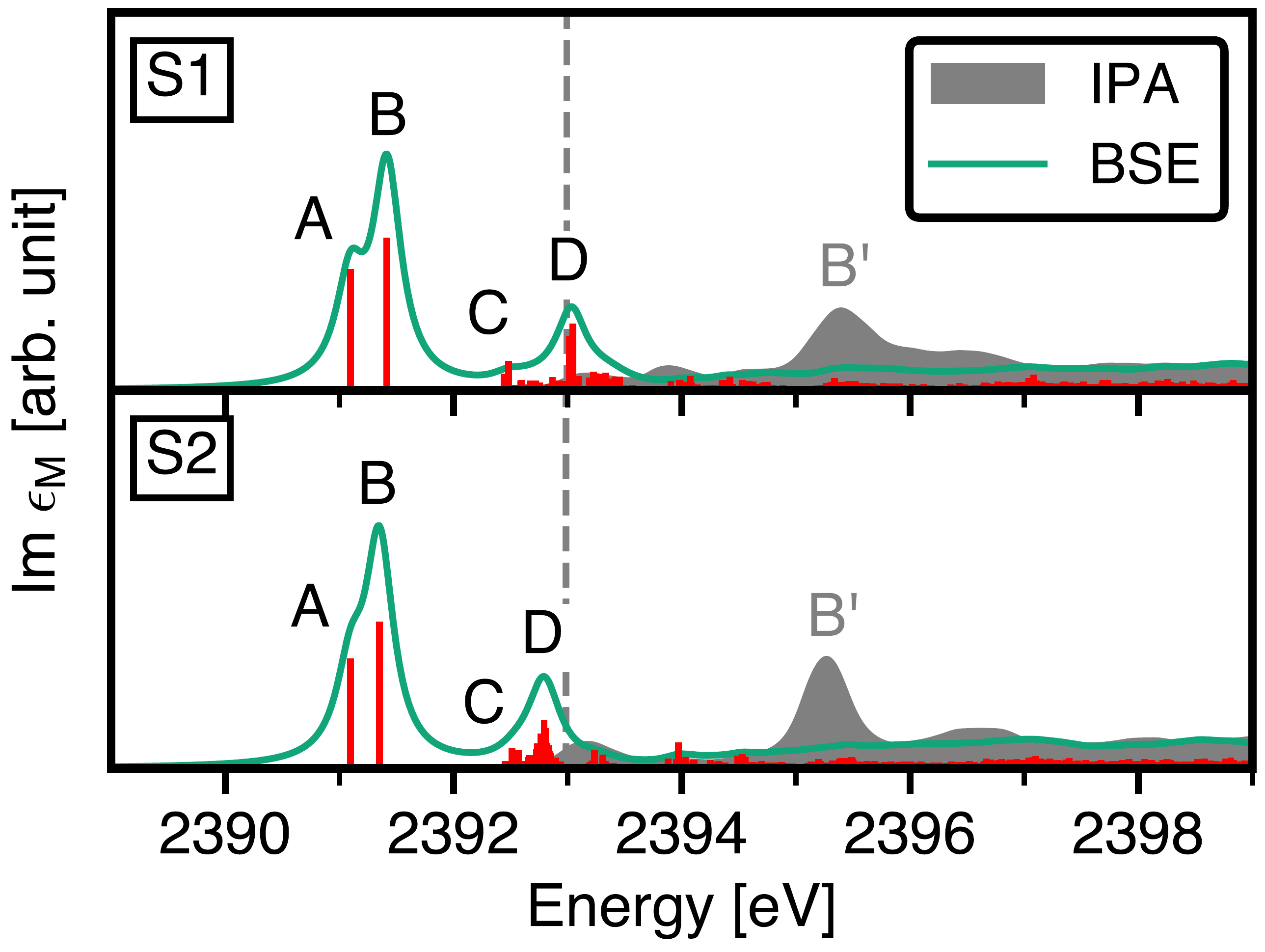}
\caption{Absorption spectra from the S $K$ edge of the inequivalent sulfur atoms in 4T (S1 and S2) averaged over the diagonal cartesian components. Excitation energies of individual excitons are indicated by the red bars. For comparison, the IPA results are shown (gray areas); the dashed bars mark the corresponding onset. \label{fig:4T_SK_comparison}}
\end{figure}

\begin{figure}
\includegraphics[width=1.\linewidth]{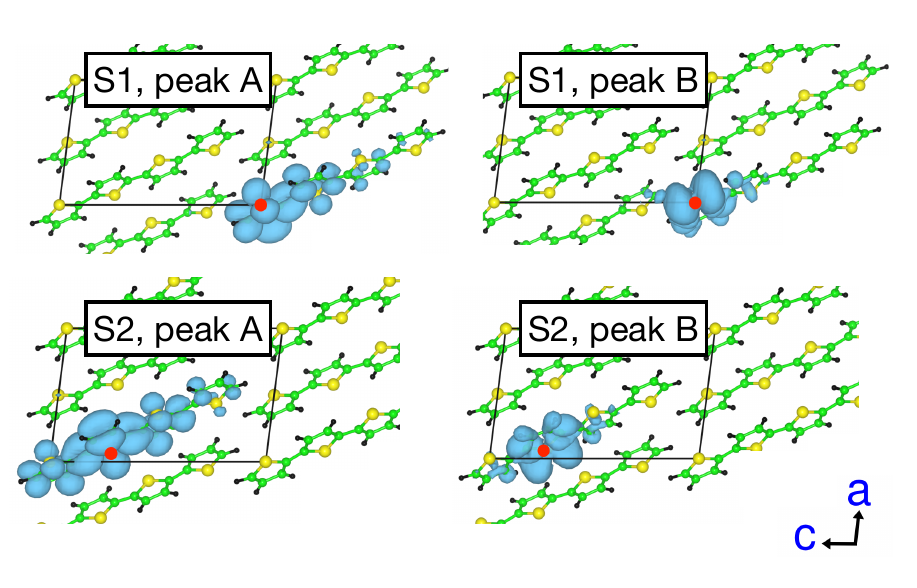}
\caption{Real-space representation of the electron distribution of the excitons in the S1 (top) and S2 (bottom) $K$ edge in crystalline 4T. The left (right) panels show the lowest $\pi^{*}$ ($\sigma^{*}$) resonances. The hole (red dot) is fixed near the probed atom. \label{fig:4T_exciton_realspace}}
\end{figure}

For the analysis of S K-edge spectra, we focus on the example of 4T, that has two inequivalent sulfur sites (labeled S1 and S2). The corresponding BSE spectra are shown in Fig.~\ref{fig:4T_SK_comparison} together with their IPA counterparts. Analogous to the C $K$ edge, we can directly relate the sulfur $p$ contributions of the conduction states (Fig.~\ref{fig:PDOS}) to the spectral features, which are dominated by an intense peak B', formed by transitions to bands with $\sigma^*$ character which are associated with the single C-S bond. This corresponds to the peak at \SI{3.0}{\electronvolt} in the PDOS. At the IPA onset, we find peaks with low intensity. They represent transitions to the LUMO+n subbands with $\pi^*$ character. In this range, excitations from S2 contribute more significantly to the LUMO subbands, whereas those from S1 contribute equally to all LUMO+n subbands. This result is expected since the LUMO is less localized on the sulfur atom at the edge of the molecule (S1). 

\begin{table}
\caption{Excitation energies, $E$, of the spectral features in the S $K$ absorption edge of crystalline 4T and assignment of features to the respective final states. The excitation energies are defined for each inequivalent atom by the most intense exciton contributing to the respective peaks. \label{tab:SK_edge_peaks}}
\begin{ruledtabular}
\begin{tabular}{lcc}
 {Peak / Assignment} & \multicolumn{2}{c}{$E$ [\si{\electronvolt}]} \\
 \cline{2-3}
 & S1 & S2 \\
 \hline \\[-8pt]
A / $\pi^*$(LUMO,LUMO+1,LUMO+2)   & 2391.1 & 2391.1  \\
B / $\sigma^*$(C-S)               & 2391.4 & 2391.4 \\
C / mixed $\sigma^*$, $\pi^*$      & 2392.4 & 2392.6 \\
D / mixed $\sigma^*$, $\pi^*$      & 2393.0 & 2392.8 \\
\end{tabular}
\end{ruledtabular}
\end{table}

The inclusion of electron-hole interaction redshifts the spectrum by \SI{1.9}{\electronvolt} and redistributes the oscillator strength to a few excitons in the vicinity of the absorption onset. For this deep edge, we find that LFE play a minuscule role, only inducing a negligible shift of less than \SI{1}{\milli\electronvolt} (see Fig.~\ref{fig:all_spectra_comparison_LFE} in the Appendix). A summary of the relevant spectral features is given in Table~\ref{tab:SK_edge_peaks}. The two inequivalent sulfur atoms, with their 1$s$ levels separated by \SI{6}{\milli\electronvolt}, contribute equally to the total absorption spectrum which is characterized by two main features. The lower one, comprising peaks A and B, separated by \SI{0.3}{\electronvolt}, is dominated by two excitons with large oscillator strengths, whereas the higher one, comprising peaks C and D, is formed by many transitions with lower intensity. The first peak, A, is formed by transitions to the subbands with $\pi^*$ character associated with the LUMO orbital. Peak B is most intense and is formed by transitions to the subbands with $\sigma^*$ character associated with the C-S bond. This is visualized in the Appendix in Fig.~\ref{fig:4T_SK_excitons}, where we show which bands contribute to these two bright excitons. Our assignment of peaks A and B matches experimental results for thiophene multilayers~\cite{Cu_thiophene} and molecular thiophene~\cite{george,perera,thiophene_CK}. 

Exciton A has a binding energy of \SI{1.90}{\electronvolt} in the spectrum from S1 and \SI{1.89}{\electronvolt} from S2. For peak B, we obtain binding energies of \SI{3.98}{\electronvolt} and \SI{3.93}{\electronvolt} from the excitation of the two species, respectively. They are, remarkably, more than twice as large as those of exciton A. (Note the importance of determining the binding energy by assessing the impact of electron-hole interaction, see Section \ref{sec:1}.) This difference can be understood by the varying degrees of localization of the e-h wavefunctions shown in Fig.~\ref{fig:4T_exciton_realspace}. While peak A is formed by transitions to $\pi^*$ orbitals that are delocalized along the oligomer chain, peak B is dominated by transitions to $\sigma^*$ orbitals, making the excitons strongly localized around the excited sulfur atoms. Moreover, we find that the wavefunction of both excitons is more delocalized for transitions from the atom in the inner thiophene ring, S2. This, in turn, leads to the smaller exciton binding energies compared to those from S1. The splitting of the excitonic states corresponding to peaks A and B are \SI{1}{\milli\electronvolt} and \SI{3}{\milli\electronvolt}, respectively. The most pronounced differences between the two inequivalent sulfur atoms occur for peaks C and D. They are formed by several transitions primarily to $\pi^*$ orbitals with some admixture of $\sigma^*$ states, in contrast to a previous assignment to transitions to higher lying $\sigma^*$ orbitals~\cite{thiophene_CK}. The energy difference in the spectra obtained from S1 and S2 is only \SI{30}{\milli\electronvolt} for peak C while it is \SI{250}{\milli\electronvolt} for D. As a result, peak C appears barely as a shoulder of peak D in the spectrum arising from S2 (see Fig.~\ref{fig:4T_SK_comparison}).

In order to analyze in more detail the S $K$-edge spectra as a function of the chain length, we go back to Fig.~\ref{fig:all_spectra_comparison}, inspecting the middle panels. The overall spectral shapes and intensities are very similar for all investigated systems. The main differences occur for the two lowest lying peaks. The first one (labeled A in Fig.~\ref{fig:4T_SK_comparison}) is blueshifted with increasing oligomer length, and like in the spectra from the C $K$ edge, this effect is stronger when going from 2T to 4T (\SI{0.28}{\electronvolt}) than from 4T to 6T (\SI{0.03}{\electronvolt}). For peak B, the corresponding energy differences are \SI{0.07}{\electronvolt} and \SI{0.08}{\electronvolt}, respectively. The $\pi^*$ resonances, on the other hand, are hardly affected by the oligomer length. A similar result has been found for $\alpha$-substituted thiophenes~\cite{george}. Analogous to the C $K$ edge, we attribute this blueshift to the reduction of the average Coulomb attraction with increasing oligomer length. Peaks A and B are separated by \SI{0.5}{\electronvolt} in 2T, \SI{0.3}{\electronvolt} in 4T, and \SI{0.3}{\electronvolt} in 6T. For comparison, experimental values of \SI{0.5}{\electronvolt} for thiophene multilayers~\cite{Cu_thiophene} and \SI{0.7}{\electronvolt} for thiophene in solution~\cite{george} have been reported. This is inline with our observations of pronounced exciton localization on short or isolated molecules. As such, we also expect a value larger than \SI{0.5}{\electronvolt} for monothiophene crystals.

In Fig.~\ref{fig:all_spectra_comparison}, middle panels, we also distinguish the contributions from inequivalent sulfur atoms. The first two peaks are nearly identical for all sulfur atoms, with differences of less than \SI{60}{\milli\electronvolt} which is the order of the S 1$s$ core-level shift  (\SI{33}{\milli\electronvolt}) in 6T. In 4T and 6T, the third peak from S1 is blueshifted by 0.2-0.3\,\si{\electronvolt} with respect to its counterparts from the other sulfur atoms. This result is somewhat suprising since we would rather expect a small redshift of the $\pi^*$ resonances of S1 due to reduced correlation effects for its position at the edge of the molecule. A possible explanation lies in the different contributions of the two sulfur atoms to the electronic structure. This will be discussed in connection to the S $L_{2,3}$ edge below. We note that a similar finding has not been reported before since most studies concentrate on thiophenic compounds with only one sulfur atom~\cite{Cu_thiophene,george,mijovilovich,thioether}. 

Lastly, we address the trends of the exciton binding energies as summarized in Fig.~\ref{fig:C_K_S_K_exciton_comparison}. Analogous to the results obtained from the C $K$ edge, the exciton binding energy of peak A is reduced by about \SI{1}{\electronvolt} when going from 2T to 6T. Overall, we find smaller values for the spectra obtained from S atoms in the inner thiophene rings ($\te{S1}>\te{S2}>\te{S3}$). The exciton binding energies of peak B are also slightly reduced with increasing molecular length, the reduction from 2T to 6T being \SI{0.4}{\electronvolt}. The smaller decrease compared to peak A is explained by the character of the transition, as $\sigma^*$ resonances are less affected by the increased aromatic character of longer oligomers and remain largely localized on the probed atom and on the respective monomer (see also Fig.~\ref{fig:4T_exciton_realspace}). 
We note in passing that to the best of our knowledge, there are no experimental references for the S $K$-edge of oligothiophene crystals except monothiophene \cite{thiophene_CK,perera,mijovilovich}. 

\subsubsection{Sulfur $L_{2,3}$ edge \label{sec:S_L23_edge}}

\begin{figure} \centering
\includegraphics[width=0.95\linewidth]{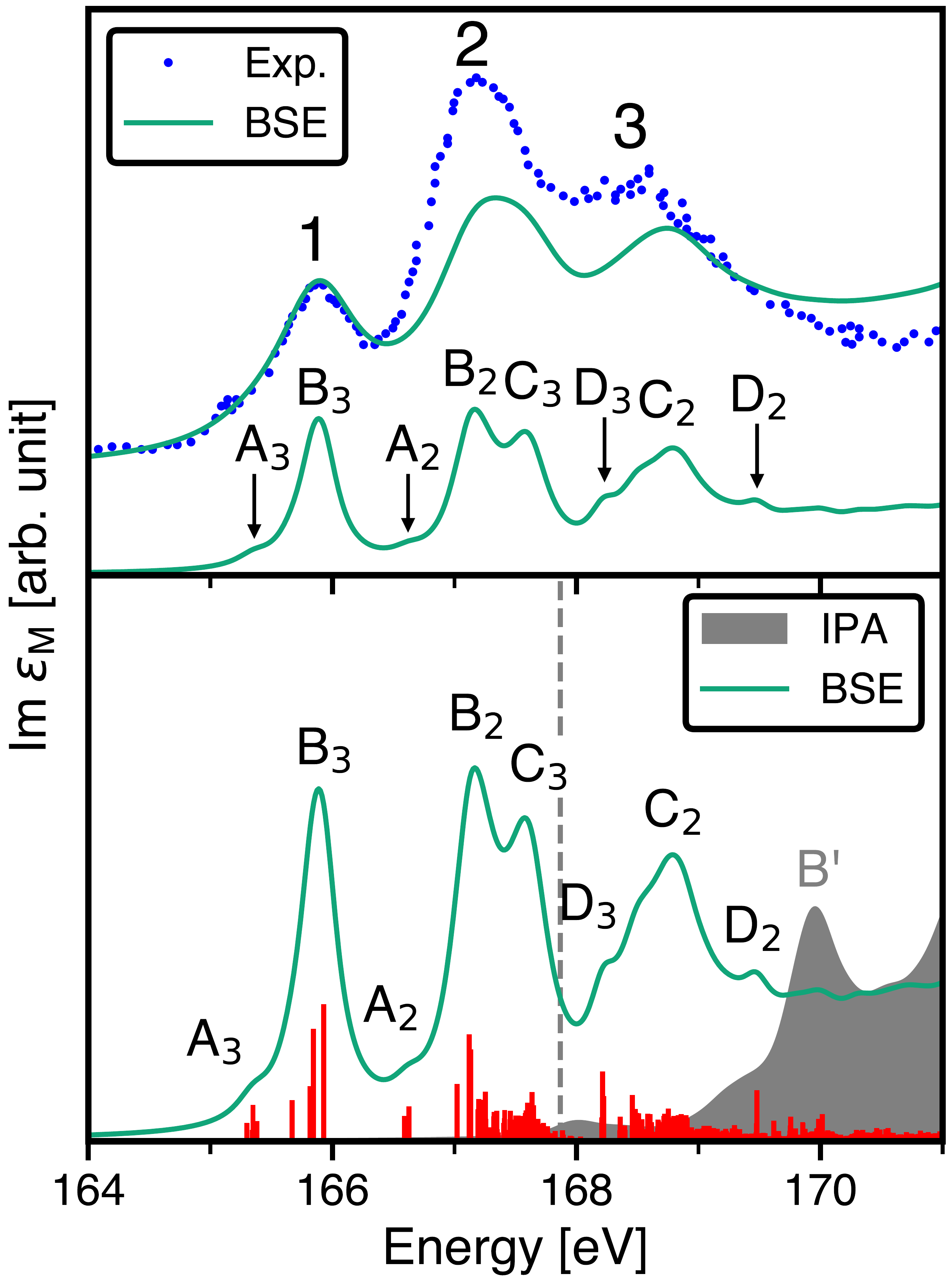}
\caption{Top: S $L_{2,3}$ absorption spectra of 2T, averaged over the diagonal cartesian components (green line) compared to experiment (blue dots) ~\cite{Agthiophene}. The calculated spectra are shifted by \SI{15.3}{\electronvolt} to align the first peak with its experimental reference. A Lorentzian broadening of \SI{400}{\milli\electronvolt} is applied in the top curve; a smaller value of \SI{150}{\milli\electronvolt} is used in the bottom curve to resolve all spectral features. Bottom: BSE spectra and excitation energies (red) with  their oscillator strength indicated by the height of the bars. A Lorentzian broadening of \SI{150}{\milli\electronvolt} is applied. For highlighting the strong excitonic effects, the IPA solution (gray area) is displayed for comparison, where the dashed line represents the absorption onset. \label{fig:2T_L23_comparison_exp}}
\end{figure}

\begin{figure}
\includegraphics[width=0.95\linewidth]{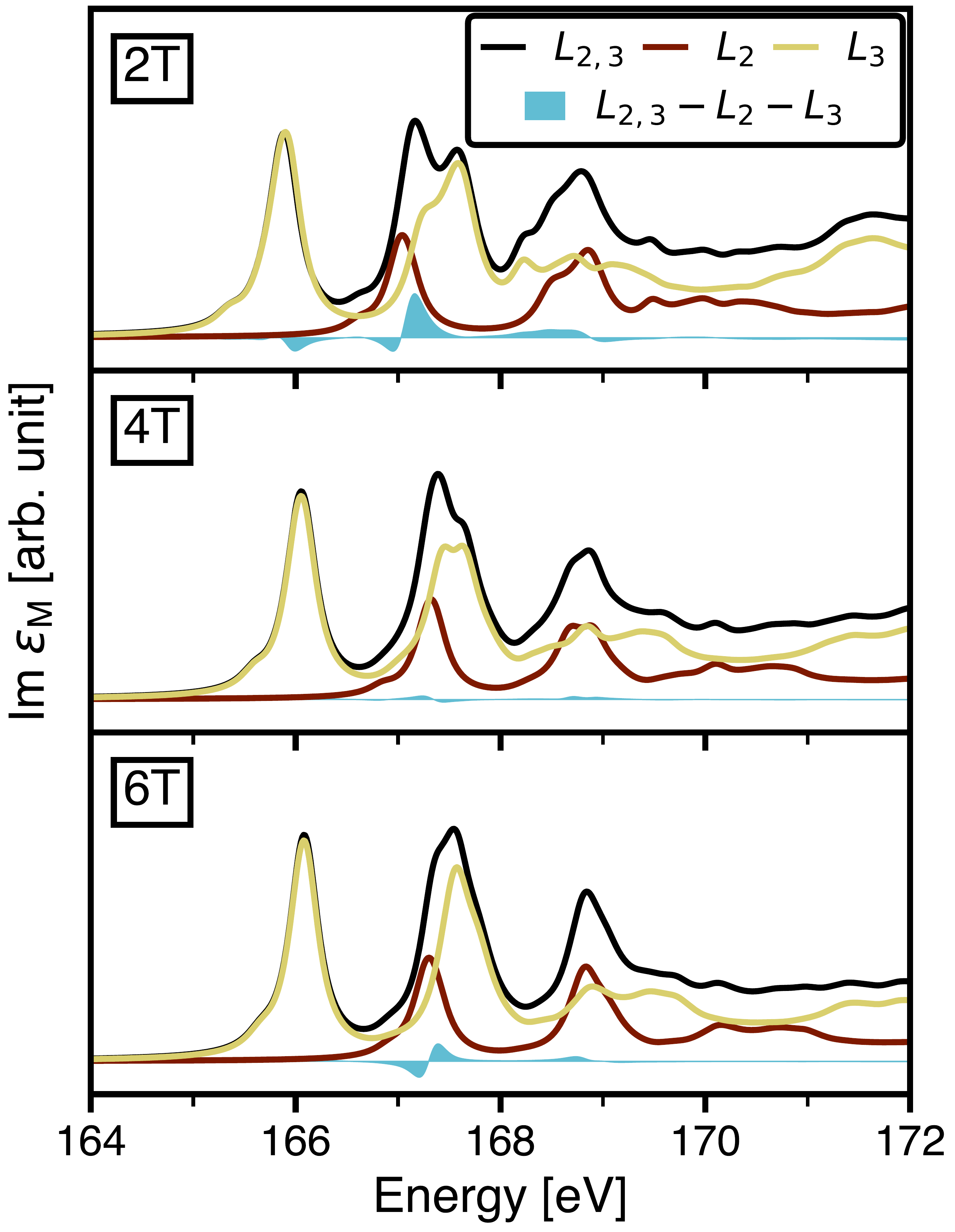}
\caption{Sulfur $L_2$ and $L_3$ BSE spectra and cross terms, $L_{2,3}-L_2-L_3$, for crystalline nT. The shown curves represent sums of the contributions from all inequivalent S atoms and averages over the three diagonal cartesian components.\label{fig:S_L2_L3_comparison}}
\end{figure}

\begin{table}
\caption{Excitation energies, $E$, of the spectral features in the S $L_{2,3}$ absorption edge of crystalline 2T, exciton binding energies, $E_b$, and assignment of features to the respective final states. The excitation energies are defined by the most intense exciton contributing to the respective peaks. The binding energies are obtained as the difference in excitation energy with respect to their IPA counterparts, \ie the IPA onset for $\pi^*$(LUMO, peak A) and peak B' for $\sigma^*$(C-S, peak B) resonances.\label{tab:SL23_edge_peaks}}
\begin{ruledtabular}
\begin{tabular}{llcl}
 {Peak}& {$E$ [\si{\electronvolt}]} & {$E_b$ [\si{\electronvolt}]} & {Assignment} \\
 \hline \\[-8pt]
A$_3$    &  {165.3} & 2.6 & {2$p_{3/2}\rightarrow\pi^*$(LUMO,LUMO+1)}   \\
B$_3$    &  {165.8} & 4.3 & {2$p_{3/2}\rightarrow\sigma^*$(C-S)}    \\
A$_2$    &  {166.6} & 1.3 & {2$p_{1/2}\rightarrow\pi^*$(LUMO,LUMO+1)}    \\
B$_2$    &  {167.0} & 3.0 & {2$p_{1/2}\rightarrow\sigma^*$(C-S)} \\
& & &  {2$p_{3/2}\rightarrow\pi^*$(LUMO)} \\
C$_3$    &  {167.6} & 0.6 &  {2$p_{3/2}\rightarrow\pi^*$(LUMO)} \\
& & &  {2$p_{1/2}\rightarrow\sigma^*$(C-S)} \\
D$_3$    &  {168.2} &  & {2$p_{3/2}\rightarrow\te{mixed }\sigma^*$, $\pi^*$}   \\
C$_2$    &  {168.5} &  & {2$p_{1/2}\rightarrow\pi^*$(LUMO)}   \\
D$_2$    &  {169.5} &  &  {2$p_{1/2}\rightarrow\te{mixed }\sigma^*$, $\pi^*$}  \\
\end{tabular}
\end{ruledtabular}
\end{table}

In Fig.~\ref{fig:2T_L23_comparison_exp}, the S $L_{2,3}$ absorption edge of 2T obtained from the BSE and the IPA is shown. Since dipole-allowed transitions occur only from the $p$-like core state to conduction states with $s$ or $d$ character, the IPA spectrum reflects the symmetry-decomposed features in the unoccupied bands. At the IPA onset, we find a hump with low intensity. It is formed by transitions to the LUMO subbands which exhibit small contributions from the S $d$ states. The intense peak at \SI{170}{\electronvolt} corresponds to the peaks at \SI{3.0}{\electronvolt} in the PDOS (see Fig.~\ref{fig:PDOS}). The solution of the BSE redshifts the spectrum by \SI{3}{\electronvolt} compared to the IPA spectrum, and the oscillator strength is redistributed to a few excitons. LFE, however, do not play a significant role in the formation of excitons and shift the excitation energies by less than \SI{150}{\milli\electronvolt}~(see Fig.~\ref{fig:all_spectra_comparison_LFE} in the Appendix).

The absorption spectra of 2T, shown in Fig.~\ref{fig:2T_L23_comparison_exp}, match very well the experimental SOC splitting of \SI{1.3}{\electronvolt} in the S $2p$ levels~\cite{spin_splitting_L23,2spin_splitting_L23}. To guide the reader, we label the spectral features in addition to denoting the edge. The three main features -- peak B$_3$, the peak structure formed by peaks B$_2$ and C$_3$, and peak C$_2$ -- are comparable in intensity. An enumeration of all spectral features and corresponding binding energies is given in Table~\ref{tab:SL23_edge_peaks}. We first analyze the features originating from the $L_3$ edge, \ie the S 2$p_{3/2}$ states. The shoulder A$_3$ is formed by four excitons with low oscillator strengths, corresponding to transitions to the LUMO and LUMO+1 subbands with $\pi^*$ character. The intensity of this feature is small due to the predominant $p$ character of the LUMO subbands. For the lowest-lying exciton, we calculate a binding energy of \SI{2.6}{\electronvolt} and an exciton splitting of \SI{44}{\milli\electronvolt}. It spreads over the respective 2T molecule, but there is no charge transfer to adjacent molecules. The second peak, B$_3$, is solely formed by transitions to the $\sigma^*$ orbitals associated with the C-S bond. It is dominated by four excitons with high oscillator strengths. They are highly localized, as evident by the larger binding energy of \SI{4.3}{\electronvolt} compared to A$_3$. The splitting of the two lowest-lying excitons of peak B$_3$ is \SI{150}{\milli\electronvolt}. Remarkably, despite S $L_{2,3}$ being a much shallower edge than S $K$, the binding energies of peaks A$_3$ and B$_3$ are almost equal to the ones found in the S $K$ edge of 2T. The exciton splittings, however, are considerably larger here, ranging from \SIrange{130}{150}{\milli\electronvolt}. They are of the same order as those found in the C $K$-edge concomitant with the comparable energies of the C 1$s$ and S 2$p$ core states.  

Several excitons contribute to the third peak, C$_3$. It is predominately formed by transitions with $\pi^*$ character to the LUMO subbands. The overlap with peak B$_2$ leads to an additional mixing with transitions from the S 2$p_{1/2}$ state to the $\sigma^*$ bands, but this does not significantly alter its excitonic character. Finally, we identify the shoulder D$_3$ at \SI{168.2}{\electronvolt} being formed by transitions with mixed $\pi^*$ and $\sigma^*$ character to the LUMO+1 subbands and higher bands up to \SI{5.0}{\electronvolt}, which are of mixed S $s$ and $d$ character. In the $L_2$ edge, the spectral features are blueshifted by \SI{1.3}{\electronvolt} due to SOC, where the transitions occur to the same conduction states as in the $L_3$ edge. Peak B$_2$ results from significant mixing between transitions from the S 2$p_{3/2}$ states to the LUMO subbands and transitions from the S 2$p_{1/2}$ states to bands with $\sigma^*$ character. This is also visualized in Fig.~\ref{fig:2T_SL23_excitons}, where we show for selected examples which bands contribute to the excitons with highest oscillator strengths.

In Fig.~\ref{fig:2T_L23_comparison_exp}, we also compare our results with the experimental spectrum obtained for 2T multilayers on Ag(111)~\cite{Agthiophene}. In lack of information on the setup of the measurement, we display the theoretical result as the average over the diagonal cartesian components of the macroscopic dielectric tensor. The three intense resonances found in the experimental spectrum are well replicated by our calculation. Note that the experimentally observed shoulder at \SI{165.3}{\electronvolt} corresponds to feature A$_3$ that becomes apparent when using a smaller broadening of \SI{150}{\milli\electronvolt} (lower curve). In excellent agreement with our results (see Table~\ref{tab:SL23_edge_peaks}), all previous studies assign resonance (1) to transitions from S 2$p_{3/2}$ states to $\sigma^*$ orbitals associated with the C-S bond~\cite{Agthiophene,Ramsey_SL23,Koller_SL23,thiophene_CK}. 
We trace back the second resonance (2) to the superposition of S~2$p_{1/2}\rightarrow\sigma^*$~(C-S) transitions and S~2$p_{3/2}\rightarrow\pi^*$~(LUMO) transitions. in contrast to Ref. ~\cite{Agthiophene} where it was assigned to a superposition of S~2$p_{1/2}\rightarrow\sigma^*$~(C-S) transitions and Rydberg-like transitions from the S~2$p_{3/2}$ states to higher energy $\sigma^*$~orbitals. The splitting between the two peaks contributing to (2), B$_2$ and C3, is \SI{0.4}{\electronvolt}, in excellent agreement with the experimental value of \SI{0.4}{\electronvolt}~\cite{Koller_SL23}. Feature (3) is formed by S~2$p_{1/2}\rightarrow\pi^*$~(LUMO) transitions in agreement with other experimental results~\cite{Ramsey_SL23,Koller_SL23,thiophene_CK}. 

Some effort has also been spent to describe the less intense peaks in the S~$L_{2,3}$ spectrum commonly observed in the experimental results~\cite{Koller_SL23,thiophene_CK,Agthiophene}. Here, we provide a comprehensive assignment of all spectral features: We attribute the shoulder A$_3$, at \SI{165.3}{\electronvolt}, observed in multiple experiments~\cite{Koller_SL23,Agthiophene}, to S~2$p_{3/2}\rightarrow\pi^*$~(LUMO, LUMO+1) transitions. Peak A$_2$ at \SI{166.6}{\electronvolt}, however, has not been resolved experimentally. Koller \textit{et al.}~\cite{Koller_SL23} found that resonance (3) is straddled by two weaker resonances, which they assign to transitions to the LUMO+1 subbands. Our calculation matches this result very well, as two weaker resonances, D$_3$ and D$_2$, straddle C$_2$, the intense one. D$_3$ and D$_2$ are formed by transitions with mixed $\pi^*$ and $\sigma^*$ character from the S~2$p_{1/2}$ and S~2$p_{3/2}$ states to the LUMO+1 subbands and higher bands up to \SI{5.0}{\electronvolt} above the onset (see also PDOS in Fig.~\ref{fig:PDOS}).

The individual $L_2$ and $L_3$ spectra together with the cross terms are displayed in Fig.~\ref{fig:S_L2_L3_comparison}. The $L_2$ spectrum is, according to SOC, blueshifted by \SI{1.3}{\electronvolt} compared to to the $L_3$ counterpart. In the independent-particle picture, we obtain, as expected, a branching ratio of $2:1$ for the $L_{3}$ and $L_{2}$ sub-edges, which reflects the ratio between the numbers of $M_{J}$-states. The cross terms significantly lower the branching ratio by transferring intensity from the $L_3$ to the $L_2$ edge. This effect is most pronounced from \SIrange{167.0}{167.5}{\electronvolt} in 2T and 6T where significant mixing of both sub-edges occurs. Similar results have been found for the $L_{2,3}$ absorption edge of 3$d$ transition elements, \eg in TiO$_2$ where the branching ratio is reduced to approximately $1:1$~\cite{vorwerk2, begum2023}.

Finally, we address the binding energies of the lowest-lying excitons in all investigated systems (Fig.~\ref{fig:C_K_S_K_exciton_comparison}). The value for peak A$_3$ is reduced by about \SI{1}{\electronvolt} when going from 2T to 6T, again owing to the exciton delocalizaton with increasing oligomer length. Smaller values are found for the spectra from the S atoms in the inner thiophene rings~($\te{S}_1>\te{S}_2>\te{S}_3$). The binding energies of the excitons corresponding to peak B$_3$ exhibit only a range of \SI{0.5}{\electronvolt} across the different systems. This small range compared to peak A$_3$ originates from the nature of the transitions as $\sigma^*$ resonances remain largely localized on the probed atom and on the respective monomer. They are, thus, less affected by the increased aromatic character of longer oligomers. These results are almost identical to the ones obtained for peaks A and B in the S $K$ edge, owing to the similar nature of the transitions. In both absorption edges, they target the same final states, which are for (1), the LUMO+n subbands and for (2), the bands of $\sigma^*$ character associated with the C-S bond at approximately \SI{3}{\electronvolt} (see PDOS in Fig.~\ref{fig:PDOS}). These bands have mainly C $p$ and S $p$ character but exhibit admixtures of S $d$ states. This hybridization is decisive for the similar exciton properties of the S $K$ and S $L_{2,3}$ absorption edges.

\section{Summary and Conclusions} 
We have presented a first-principles study of core excitations in oligothiophene crystals, \ie 2T, 4T, and 6T, treating the absorption from the $K$ and $L_{2,3}$ edges on the same footing. In all spectra, we have found that the inclusion of electron-hole interaction leads to a significant redshift of the absorption onset up to \SI{3}{\electronvolt}. At all edges, several bound excitons with binding energies of up to \SI{4.5}{\electronvolt} are formed. Their final states exhibit the $\pi^*$ orbital character of the lower-lying conduction bands. However, excitations with $\sigma^*$ character have the largest binding energies. The overall spectral shape and intensity of the main peaks in all considered absorption edges is very similar in all investigated systems. The exciton binding energies, however, are decreasing by up to \SI{1.0}{\electronvolt} going from 2T to 6T. This results from the reduction of the average Coulomb attraction, due to increased delocalization of the e-h pairs with increasing oligomer length together with enhanced dielectric screening. $\pi^*$ resonances, which are delocalized along the molecular chain, are affected more strongly than $\sigma^*$ resonances, which are localized on the respective excited atoms. In the absorption from the C $K$-edge, spectral features can be assigned to two groups of carbon atoms, \ie with or without sulfur bonding. The differences among inequivalent sulfur atoms are much less pronounced in the absorption from the S $K$- and S $L_{2,3}$-edges. Our results for the C $K$- and S $L_{2,3}$-edges for crystalline 2T matches the experimental spectra for 2T multilayers~\cite{Agthiophene}, which highlights the predominant molecular character of the spectral features.

This comprehensive study of core excitations in oligothiophene crystals provides an in-depth characterization of these materials in terms of light-matter interaction in the short-wavelength range. Our work further confirms the predictive power of many-body perturbation theory in determining the character of the excitonic resonances and their dependence on the oligomer length.

\section{Acknowledgements}
Partial financial support by the German Research Foundation (DFG) through the Collaborative Research Centers 658 (project number 12489635) and 951 (project number 182087777) is appreciated. 

\appendix
\counterwithin{figure}{section}
\counterwithin{table}{section}

\section{Computational parameters}
For completeness, we show the employed computational settings for the different absorption edges in Tab.~\ref{tab:comp_settings}.
\begin{table}
\caption{\label{tab:comp_settings}Computational settings employed for the calculation of the absorption edges of nT crystals. Shown are the $k$-grid, the planewave cutoff $R_{\te{MT}}^{\te{min}}\,G_{\te{max}}$, the cutoff for local-field effects $\left| \bm{G}+\bm{q} \right|_{\te{max}}$ (in units of \si{\bohr^{-1}}), and the number of unoccupied states in the BSE Hamiltonian.}
\begin{ruledtabular}
\begin{tabular}{*{4}{c}}
Carbon $K$-edge & \mc{2T} & \mc{4T} & \mc{6T} \vspace{2pt}\\
\hline \\[-8pt]
$k$-grid & $6\times 6\times 4$ &  $3\times 5\times 2$ &  $8\times 8\times 6$ \vspace{2pt} \\
$R_{\te{MT}}^{\te{min}}\,G_{\te{max}}$ & 4.5 &  4.0 &  4.0 \vspace{2pt} \\
$\left| \bm{G}+\bm{q} \right|_{\te{max}}$ & 4.5 &  3.0 &  3.0 \vspace{2pt} \\
Unoccupied states & 50 &  70 &  60 \vspace{2pt} \\
\vspace{-5pt}\\
Sulfur $K$-edge & \mc{2T} & \mc{4T} & \mc{6T} \vspace{2pt}\\
\hline \\[-8pt]
$k$-grid & $6\times 6\times 4$ &  $3\times 5\times 2$ &  $3\times 5\times 2$ \vspace{2pt} \\
$R_{\te{MT}}^{\te{min}}\,G_{\te{max}}$ & 4.5 &  4.0 &  4.0 \vspace{2pt} \\
$\left| \bm{G}+\bm{q} \right|_{\te{max}}$ & 4.5 &  3.5 &  3.5 \vspace{2pt} \\
Unoccupied states & 100 &  120 &  120 \vspace{2pt} \\
\vspace{-5pt}\\
Sulfur $L_{2,3}$-edge & \mc{2T} & \mc{4T} & \mc{6T} \vspace{2pt}\\
\hline \\[-8pt]
$k$-grid & $4\times 4\times 4$ &  $3\times 5\times 2$ &  $3\times 5\times 2$ \vspace{2pt} \\
$R_{\te{MT}}^{\te{min}}\,G_{\te{max}}$ & 4.5 &  4.0 &  4.0 \vspace{2pt} \\
$\left| \bm{G}+\bm{q} \right|_{\te{max}}$ & 5.0 &  3.0 &  3.0 \vspace{2pt} \\
Unoccupied states & 100 &  100 &  130 \vspace{2pt} \\
\vspace{-5pt}\\
\end{tabular}
\end{ruledtabular}
\end{table}

\section{X-ray absorption spectra}

\begin{figure*}[h!]
\includegraphics[width=1.0\linewidth]{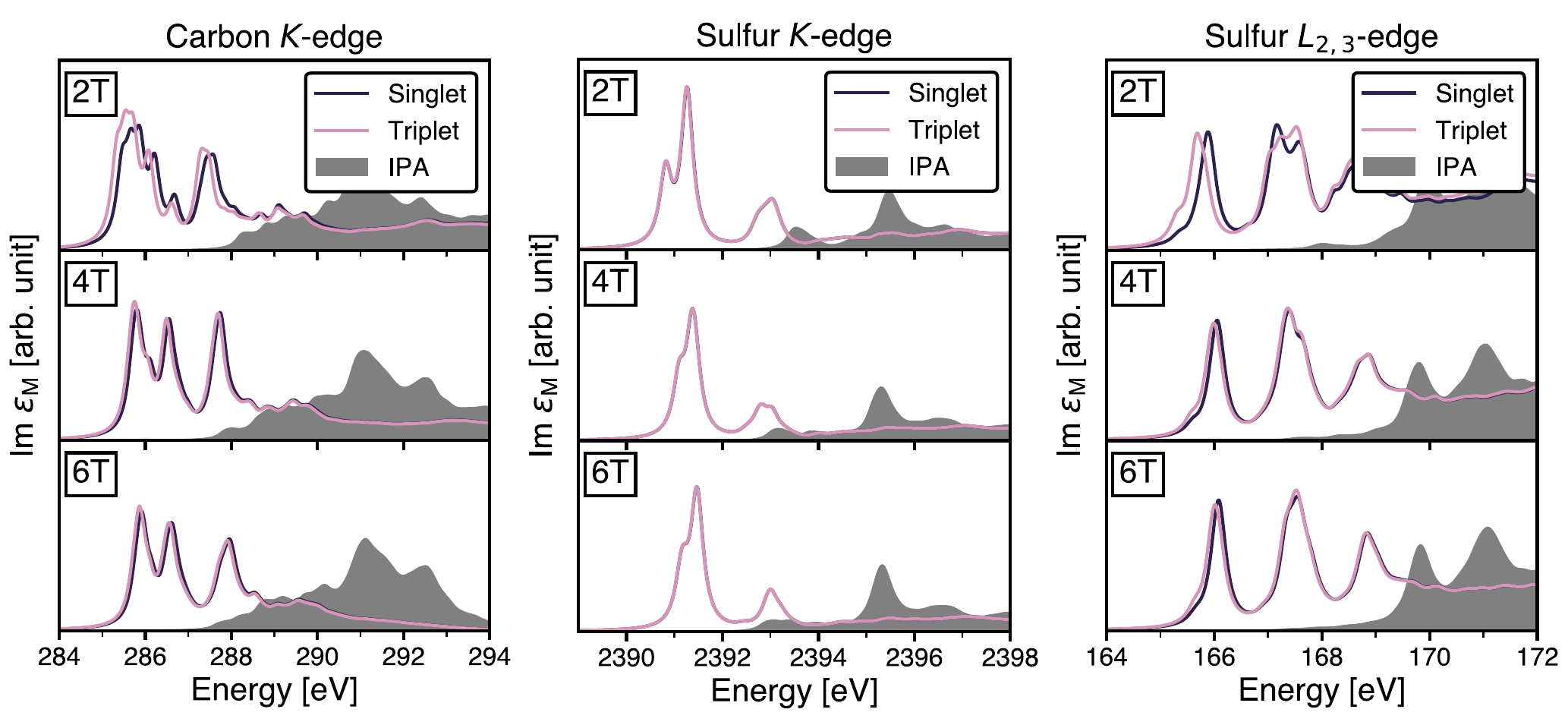}
\caption{Absorption spectra from the carbon $K$ (left), sulfur $K$ (middle), and sulfur $L_{2,3}$ (right) edge of crystalline 2T~(top), 4T~(middle), and 6T~(bottom). The shown curves represent sums of the contributions from all inequivalent S atoms and averages over the three diagonal cartesian components. Shown are the singlet, the triplet~($H^x=0$), and the IPA~($H^x$,$H^c=0$) solutions to the BSE (Eq.~\eqref{eq:BSE_parts}). \label{fig:all_spectra_comparison_LFE} }
\end{figure*}

\begin{figure*}[h!]
\includegraphics[width=1.0\linewidth]{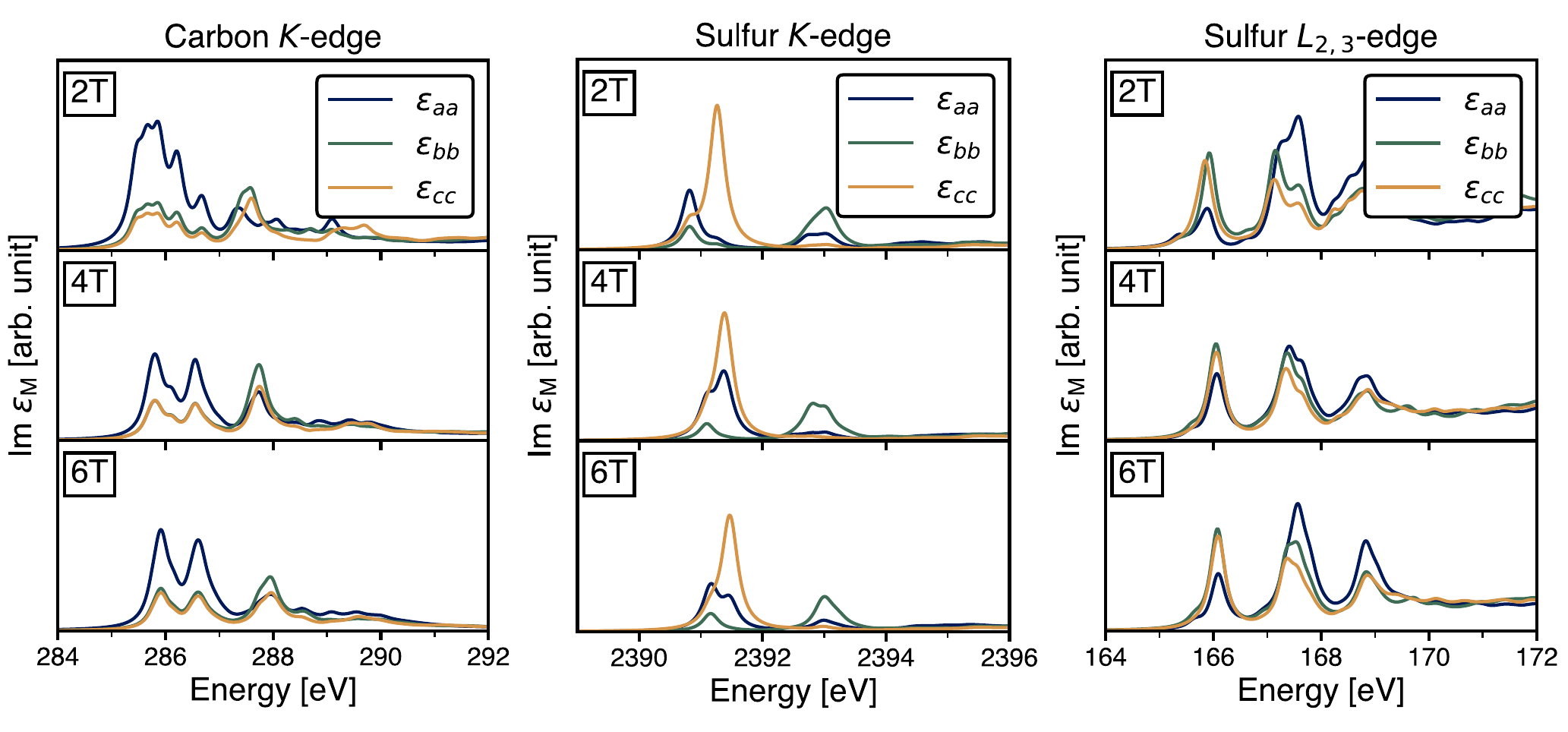}
\caption{Imaginary part of the macroscopic dielectric tensor components along the crystal axes for the C $K$ (left), S $K$ (middle), and S $L_{2,3}$ (right) edge of crystalline 2T~(top), 4T~(middle), and 6T~(bottom). They are summations over all respective inequivalent atoms.\label{fig:all_spectra_comparison_crystalline} }
\end{figure*}

\begin{figure}[h!]
\includegraphics[width=0.98\linewidth]{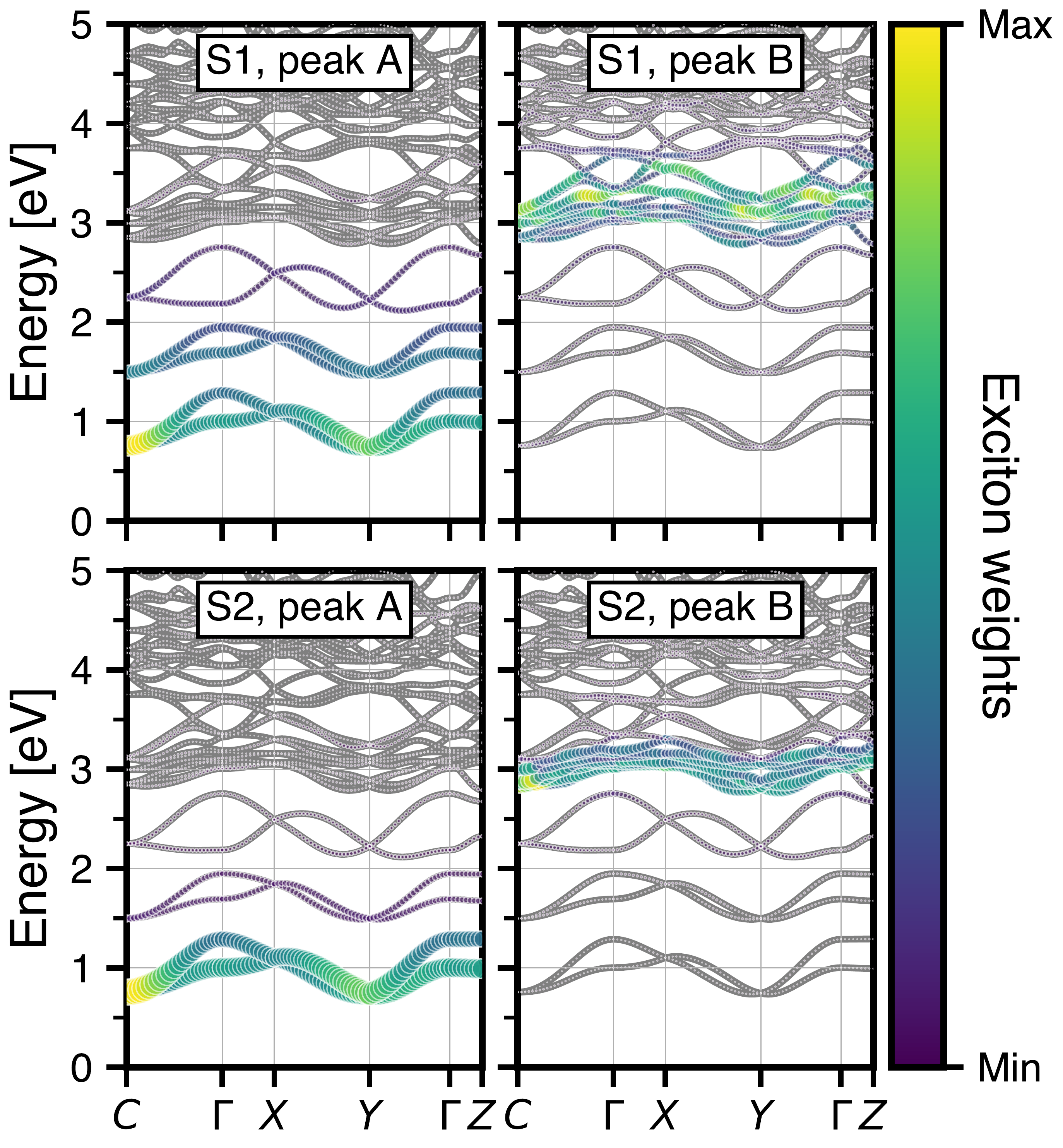}
\caption{Conduction band contributions to the bound excitons with the largest oscillator strengths at the sulfur $K$ edge of 4T. The size and color of the circles are indicative of the exciton weights. \label{fig:4T_SK_excitons} }
\end{figure}

\begin{figure}[h!]
\includegraphics[width=0.95\linewidth]{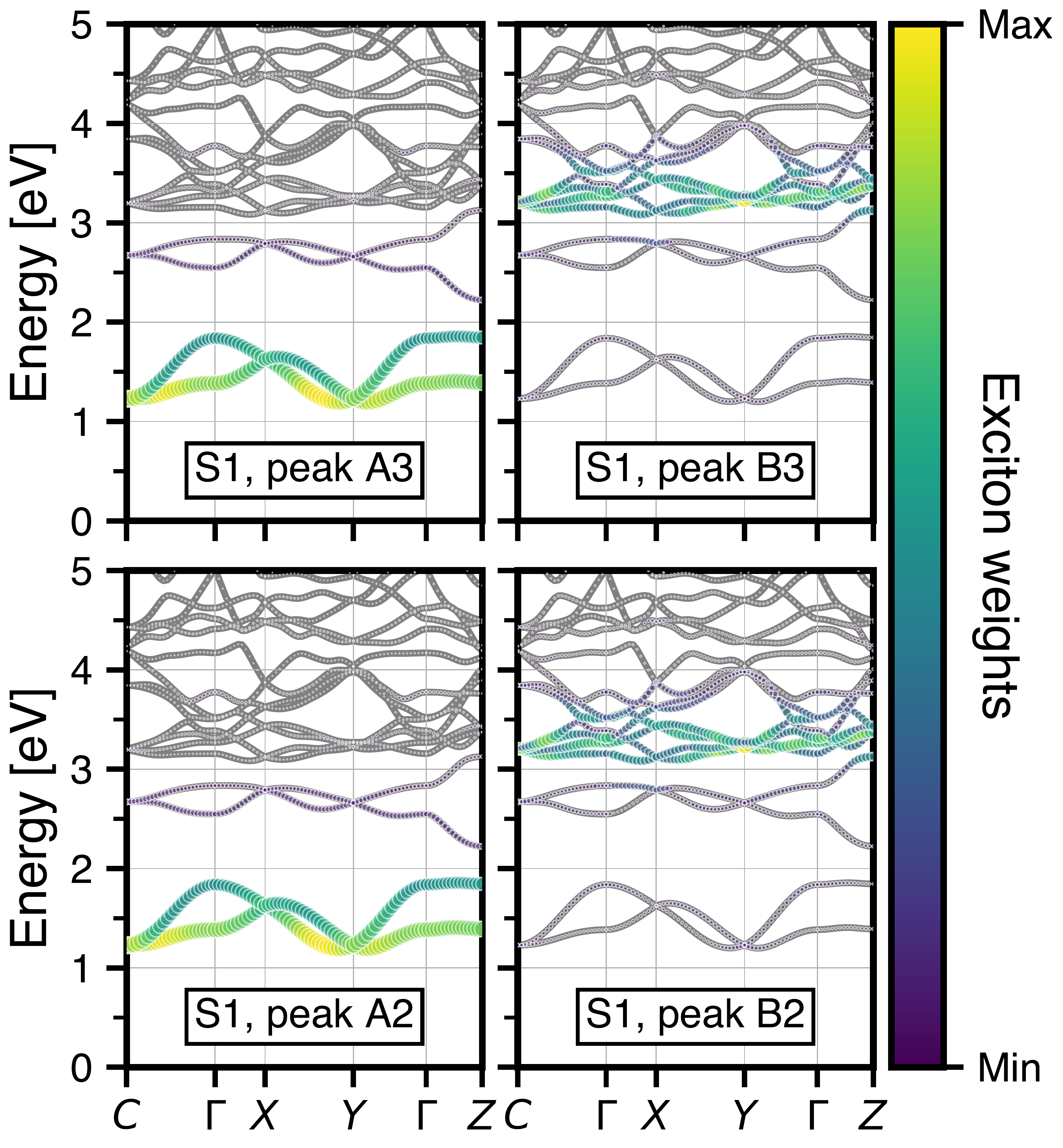}
\caption{Conduction band contributions to the bound excitons with the largest oscillator strengths at the sulfur $L_{2,3}$ edge of 2T. The size and color of the circles are indicative of the exciton weights.\label{fig:2T_SL23_excitons} }
\end{figure}

In Fig.~\ref{fig:all_spectra_comparison_LFE}, we visualize the difference between singlet and triplet excitations, \ie the impact of local-field effects, for all systems under investigation. Fig.~\ref{fig:all_spectra_comparison_crystalline} shows the individual components of the macroscopic dielectric tensors along the crystal axes. To further complement the exciton analysis in Sections \ref{sec:S_K_edge} and \ref{sec:S_L23_edge}, we depict the exciton distribution in reciprocal space for selected bound excitons in Figs. \ref{fig:4T_SK_excitons} and \ref{fig:2T_SL23_excitons}.

\section{Core-level energies}
Core-level energies of all investigated systems obtained by the local-density approximation are given in Tables \ref{tab:coreenergies_C1s} and \ref{tab:coreenergies_S1s_S2p}.
\begin{table*}[h!]
\begin{ruledtabular}
\centering
\caption{Carbon $1s$ core-level energies of crystalline nT calculated within the LDA. All energies are in units of \si{\electronvolt}.\label{tab:coreenergies_C1s}}
\begin{tabularx}{\linewidth}{@{\extracolsep{\fill} }l*{13}{c}@{}}
 & {C1} & {C2} & {C3} & {C4} & {C5} & {C6} & {C7} & {C8} & {C9} & {C10} & {C11} & {C12} \\
\hline \\[-8pt]
2T    & -261.08 & -260.54 & -260.50 & -261.50 & & & & & & & &     \\
4T   & -260.93 & -260.38 & -260.35 & -261.35 & -261.37 & -260.37 & -260.37 & -261.39 & & & &     \\
6T   & -260.75 & -260.19 & -260.16 & -261.13 & -261.14 & -260.18 & -260.17 & -261.16 & -261.16 & -260.19 & -260.19 & -261.15   \\
\end{tabularx}
\end{ruledtabular}
\end{table*}

\begin{table*}[h!]
\centering
\caption{Sulfur $1s$ and $2p$ core-level energies of crystalline nT calculated within the LDA. All energies are in units of \si{\electronvolt}.\label{tab:coreenergies_S1s_S2p}}
\begin{ruledtabular}
\begin{tabularx}{\linewidth}{@{\extracolsep{\fill} }l*{10}{c}@{}}
& \multicolumn{3}{c}{$1s$} & \multicolumn{3}{c}{$2p_{1/2}$} & \multicolumn{3}{c}{$2p_{3/2}$} \\
\cmidrule{2-4} \cmidrule{5-7} \cmidrule{8-10}
& {S1} & {S2} & {S3} & {S1} & {S2} & {S3}& {S1} & {S2} & {S3} \\
\hline \\[-8pt]
2T    & -2389.94 & & & -150.38 &  & & -149.12 & &    \\
4T   &  -2389.78 & -2389.77 & & -150.22 & -150.21 &  & -148.96 & -148.95 &   \\
6T   &  -2389.60 & -2389.59  &  -2389.56 & -150.05 & -150.04  & -150.02 & -148.79 & -148.78 & -148.76  \\
\end{tabularx}
\end{ruledtabular}
\end{table*}

%

\end{document}